\documentclass[12pt]{article}

\usepackage[utf8]{inputenc}
\usepackage[T1]{fontenc}

\usepackage[a4paper,left=2cm,right=2cm,top=2cm,bottom=2cm]{geometry}

\usepackage{graphicx}
\usepackage{textcomp}
\usepackage{amsmath}
\usepackage{amssymb}
\usepackage{mathrsfs}
\usepackage{bm}
\usepackage[authoryear]{natbib}

\usepackage{lineno}
\usepackage{placeins}
\usepackage{float}
\usepackage{longtable}
\usepackage{pdfpages}
\usepackage{subcaption}
\usepackage{titlesec}
\usepackage{appendix}
\usepackage{url}
\usepackage{comment}
\usepackage{xcolor}
\usepackage{longtable}
\usepackage{comment}

\usepackage{xcolor}
\usepackage[
    colorlinks=true,
    linkcolor=blue,
    citecolor=blue,
    urlcolor=blue
]{hyperref}

\title{\bfseries Linear and Nonlinear Latent-Space Reduced-Order Models for the\\Rayleigh--Taylor Instability}
\author{
Téo Granger$^{1,2}$,
Balu Nadiga$^{3}$,
Benoît-Joseph Gréa$^{1,2}$\thanks{Corresponding author: \href{mailto:benoit-joseph.grea@cea.fr}{benoit-joseph.grea@cea.fr}},\\
Antoine Briard$^{1,2}$,
Paul Creusy$^{1,2}$\\[1em]
\small $^{1}$CEA, DAM, DIF, F-91297 Arpajon, France\\
\small $^{2}$Université Paris-Saclay, CEA, LMCE, F-91680 Bruyères-le-Châtel, France\\
\small $^{3}$Los Alamos National Laboratory, Los Alamos, NM, USA
}

\begin{document}

\maketitle


\begin{abstract}
We use a large database of direct numerical simulations to investigate the transition of the Rayleigh--Taylor instability to turbulence and its evolution toward a late-time self-similar regime. In addition to tracking the growth of the mixing layer through the mean heavy-fluid concentration profile, we analyze one-dimensional profiles of turbulent kinetic energy and dissipation, two key quantities in classical turbulent-mixing models. We consider two reduced-order modeling strategies that differ in where nonlinearity is introduced: either in the construction of the latent space or in the description of its temporal evolution. The first method uses a linear encoder--decoder obtained using Proper Orthogonal Decomposition (POD), with nonlinear reduced dynamics learned by a physics-informed neural network (PINN). The second uses a nonlinear encoder--decoder learned by an autoencoder, while constraining the latent dynamics to remain linear and satisfy physical constraints. Both approaches achieve satisfactory performance in reconstructing, interpolating, and extrapolating the dynamics of the Rayleigh--Taylor instability.
\end{abstract}


\section{Introduction}\label{sec1}

The Rayleigh--Taylor instability (RTI) \citep{Rayleigh_1882,Taylor_1950}, which develops when a heavier fluid lies above a lighter fluid in a gravitational field, is an efficient mechanism for converting potential energy into turbulent mixing \citep{Zhou2017a,Zhou2017b}. In the context of inertial confinement fusion (ICF), this instability develops at the interface between the ablator and the fuel, leading to mixing that causes additional cooling and weakens fusion reactions \citep{Lindl_1995,Betti_2016,Remington_2019,Zhou_2025}.

Direct numerical simulations of turbulent mixing flows remain scarce and costly in many applications, including the Rayleigh--Taylor instability, despite ongoing advances in computational resources. This forces the engineering community, and in particular researchers developing predictive models for ICF, to rely on more computationally efficient approaches, such as Reynolds-Averaged Navier--Stokes (RANS) models.

Developing a consistent RANS model for turbulent flows driven by RTI remains a formidable task, not only because the large scales of RTI are highly anisotropic, but also because the flow is strongly unsteady, transitioning from a laminar to a turbulent regime while retaining some memory of the initial conditions.
These features challenge classical RANS assumptions, such as spectral equilibrium within the turbulent cascade or the Boussinesq approximation for the Reynolds stress tensor. This motivates the development of surrogate and reduced-order models (ROMs) capable of extracting the essential dynamics from high-fidelity simulations while remaining computationally affordable.\\


Beyond their use as computational accelerators, reduced-order and surrogate models have become essential tools for the analysis and modeling of complex fluid systems. Their low computational cost enables large-scale parameter explorations, uncertainty quantification, flow control, and optimization studies that would be prohibitively expensive with high-fidelity simulations alone \citep{Noack_2011}. In turbulence modeling, surrogates also provide a convenient framework for data assimilation, model calibration, and the development of improved closure models by disentangling parametric and structural modeling errors \citep{Creusy_2026}. Such applications have motivated the development of reduced representations capable of accurately reproducing the essential dynamics of high-dimensional systems while remaining computationally tractable.

Many nonlinear dynamical systems encountered in fluid mechanics involve an extremely large number of degrees of freedom. Nevertheless, their evolution is often effectively confined to a low-dimensional manifold. Reduced-order models exploit this property by seeking compact representations that retain the essential features of the underlying high-dimensional dynamics while drastically reducing the computational complexity. The construction of an efficient ROM therefore relies on two fundamental ingredients: the identification of a suitable low-dimensional representation of the flow structures and the development of an accurate model for the dynamics evolving on this reduced manifold.

Modal decomposition techniques provide a natural framework to address the first of these challenges. By separating spatial and temporal dependencies, they represent the flow field as a superposition of modes weighted by time-dependent amplitudes. Such approaches have become central tools in fluid mechanics for identifying coherent structures, analyzing nonlinear dynamics, and constructing low-dimensional models \citep{Taira_2017}. Among them, the Proper Orthogonal Decomposition (POD), introduced through the method of snapshots by Sirovich \citep{Sirovich_1987} and extensively reviewed by Berkooz et al. \citep{Berkooz_1993}, remains one of the most widely used reduction techniques owing to its optimality in representing a dataset with a low-dimensional linear basis.

The second challenge consists in modeling the evolution of the reduced variables. A classical strategy relies on Galerkin projection of the governing equations onto the POD basis, yielding a dynamical system for the modal amplitudes. However, the accuracy of such models relies on the adequacy of the reduced basis to describe the underlying dynamics. Additional modeling strategies may therefore be required, as illustrated by \citet{Noack_2003}, who introduced a shift mode to improve the representation of transient dynamics in the periodic cylinder wake by accounting for the displacement between the steady solution and the averaged flow.

Despite their success, classical POD-Galerkin models suffer from several well-known limitations. First, the POD basis is constructed from a specific ensemble of snapshots and therefore defines a reduced space associated with a particular dynamical regime and set of physical parameters. As a consequence, the POD eigenspace itself may change when control parameters are varied, reducing the predictive capability of the resulting ROM. This lack of robustness away from the reference configuration was already emphasized by \citep{Deane_1991}. Such a dependence on the training conditions becomes particularly restrictive for parametric studies, where the objective is to interpolate or extrapolate the system dynamics over a broad range of physical parameters.

Second, the assumption that a fixed set of modes can efficiently represent the flow becomes questionable for transport-dominated phenomena \citep{Rowley_2000}. Because the spatial modes are fixed, advected or moving coherent structures often require a large number of modes to be represented accurately, thereby reducing the efficiency of linear modal decompositions. This issue has motivated the development of alternative nonlinear representations and machine-learning-based mappings capable of representing transport-dominated flows more efficiently \citep{Kim_2020}.

Third, the truncation of the governing equations introduces unresolved interactions between retained and discarded modes, raising important closure and stability issues. The long-time behavior of truncated POD-Galerkin models remains a challenging issue, as emphasized in \citep{Ma_2002}. More generally, accurately reproducing the energy transfer between resolved and unresolved scales remains one of the major challenges of projection-based ROMs and requires suitable closure strategies \citep{Callaham_2023}.

Recent advances in machine learning have opened new perspectives for overcoming these limitations by learning reduced representations and reduced dynamics directly from data \citep{Brunton_2020}. Deep-learning-based approaches have been proposed to improve closure modeling and long-time predictive capabilities of reduced-order models. For example, the Complemented Deep Reduced-Order Model (CD-ROM) combines projection-based techniques with neural networks to represent unresolved dynamics and improve predictive robustness \citep{MENIER_2023_CDROM}. Similarly, physics-informed neural networks \citep{Raissi_2019} have recently been employed to construct surrogate models that enforce governing equations and physical constraints while maintaining low computational costs \citep{Thevenin_2025_JFM,Volk_2025}.

Machine learning has also enabled the development of nonlinear latent representations through autoencoders. From a theoretical perspective, linear autoencoders are closely related to POD and recover the same optimal subspace under appropriate assumptions \citep{Bourlard_1988,Bourlard_2022, Bousquet_2026}. More generally, autoencoders may be interpreted as nonlinear extensions of POD in which the data are projected onto a nonlinear manifold rather than onto a linear subspace \citep{Bousquet_2025}. Such embeddings provide an accurate framework for representing strongly nonlinear dynamics \citep{Champion_2019,Messenger_2025}. Combined with latent-space dynamics identification techniques (LasDI), they further enable the construction of parametric surrogate models by learning the evolution equations directly in the nonlinear latent space \citep{Bonneville2024, Farenga2025}.

The combination of encoders with dynamical systems identification has led to a new generation of reduced-order models. Neural networks have been employed to correct truncation errors within POD subspaces while simultaneously learning additional latent dynamics \citep{Lepage_2025}. More generally, the Koopman operator framework \citep{Koopman_1931,Koopman_1932}, provides a powerful perspective by seeking approximately linear representations of nonlinear dynamics in appropriately chosen observables. Recent works have combined autoencoders and Koopman theory to identify finite-dimensional latent spaces in which nonlinear systems evolve approximately linearly \citep{Lusch2018}. Similarly, \citep{Champion_2019} proposed to identify governing ordinary differential equations directly in the latent space of an autoencoder, although the resulting dynamics are not parameterized with respect to physical control parameters.

Finally, recent developments in scientific machine learning increasingly blur the distinction between purely physics-based and purely data-driven approaches \citep{Vinuesa_2026}. Hybrid strategies that combine physical knowledge with data-driven representations offer promising avenues to construct compact, interpretable, and robust surrogates for complex dynamical systems.\\

The present work uses this hybrid approach. We compare two reduced-order modeling strategies for the evolution of the one-dimensional profiles of a Rayleigh--Taylor flow that transitions to turbulence. Both strategies share the same overall pipeline: an encoding-decoding step that reduces the profiles to a low-dimensional representation, followed by a model of the temporal evolution of that representation. They differ in which of the two stages is allowed to be nonlinear. The first strategy encodes and decodes the profiles linearly, through a Proper Orthogonal Decomposition basis. It then learns nonlinear reduced dynamics with a physics-informed neural network (PINN). The second strategy reverses this choice. The profiles are encoded and decoded nonlinearly, by an autoencoder. The reduced-order dynamics are instead constrained to remain linear, and to satisfy physics-informed constraints. Modeling dynamics in a reduced-order latent space has been standard practice for many decades. This approach is now commonly referred to as Latent Space Dynamics Identification (LaSDI), {\citep{Bonneville2024,Farenga2025}}. Under this terminology, both strategies we consider can be seen as particular realizations of LaSDI.

\section{DNS dataset for the Rayleigh-Taylor instability}
\label{sec:theoretical_framework}

In this section, we describe the theoretical framework underlying the RT database, define the key quantities that we aim to reproduce, and provide insight into the physical phenomena observed in the simulations.
The dataset comprises $484$ direct numerical simulations spanning a broad range of initial interface perturbations, as introduced by \citet{Thevenin_2025_JFM,Thevenin_2025_Physica}. The simulations were performed on a grid of $1024^2\times 2048$ points using the in-house pseudo-spectral code \textsf{STRATOSPEC} \citep{Grea2019,Briard2020,Briard2025,Briard2025b}.


\subsection{Governing equations and nondimensionalization}

The present study is conducted within the Boussinesq approximation, which is valid for a small Atwood number $\mathcal A \ll 1$, where $\mathcal A$ quantifies the relative density contrast between the two fluids.

The flow is described by the velocity field
$\boldsymbol{U}(\boldsymbol{x},t)
=(U_x,U_y,U_z)^{\mathsf{T}}$
and the heavy-fluid concentration field $C(\boldsymbol{x},t)$. Under these assumptions, their evolution is governed by the incompressible Navier--Stokes and advection--diffusion equations:
\begin{subequations}
\label{eq:boussinesq_system}
\begin{align}
    \partial_i U_i
    &=
    0,
    \label{eq_compressib}
    \\
    \partial_t U_i
    +
    U_j\partial_j U_i
    &=
    -\partial_i P
    -
    2\mathcal{A} g \delta_{iz} C
    +
    \nu\,\partial_{jj}U_i,
    \label{eq_boussi}
    \\
    \partial_t C
    +
    U_i\partial_i C
    &=
    \kappa\,\partial_{jj}C,
    \label{eq_diffusion}
\end{align}
\end{subequations}
where repeated indices are summed over and $\delta_{iz}$ denotes the Kronecker delta. Equation~\eqref{eq_compressib} enforces incompressibility; Eq.~\eqref{eq_boussi} expresses momentum conservation; and Eq.~\eqref{eq_diffusion} governs the advection--diffusion of the concentration field. Here, $P$ denotes the reduced pressure, $g$ is the magnitude of the gravitational acceleration directed downward along the $z$-axis, and $\nu$ and $\kappa$ are the constant kinematic viscosity and molecular diffusivity, respectively. In all simulations, $\nu=\kappa$, corresponding to a Schmidt number $\mathrm{Sc}=\nu/\kappa=1$.

Physical quantities are rendered dimensionless using the characteristic length $\ell_c$ and time $t_c$, constructed from the kinematic viscosity $\nu$ and the reduced acceleration $\mathcal{A}g$:
\begin{align}
    \ell_c
    &=
    \left(
    \frac{\nu^2}{\mathcal{A}g}
    \right)^{1/3},
    &
    t_c
    &=
    \left(
    \frac{\nu}{(\mathcal{A}g)^2}
    \right)^{1/3}.
    \label{eq:characteristic_scales}
\end{align}
Hereafter, the quantities are considered dimensionless. This nondimensionalization is particularly important because the surrogate models presented in this work are intended to predict physical quantities across a broad range of conditions, thereby promoting generalizability.

\subsection{Initial setup}

A complete description of the initial setup used in the simulations is provided by \citet{Thevenin_2025_JFM,TheveninDatabase_2025}. The heavy fluid is initially placed above the lighter fluid along the vertical $z$-direction, with the two fluids separated by a diffuse interface. The initial velocity field is zero, and the interface is subjected to a random perturbation. The two-dimensional spectrum of this perturbation is characterized by its mean wavenumber, spectral bandwidth, and mean amplitude. The initial condition can therefore be summarized by four dimensionless parameters: the initial Reynolds number $\mathsf{R}$, which characterizes the mean wavelength of the perturbation and the buoyancy strength; the normalized spectral bandwidth $\mathsf{B}$; the mean interface steepness $\mathsf{S}$; and the normalized thickness $\mathsf{D}$ of the unperturbed diffuse interface.

\subsection{Mean fields and turbulent 1D quantities}

In the RT simulations, the flow is statistically homogeneous within each horizontal $x$--$y$ plane. The instantaneous concentration and velocity fields can therefore be decomposed into their Reynolds averages and corresponding fluctuations as
\begin{subequations}
\begin{align}
    C(\boldsymbol{x},t)
    &=
    \overline{C}(z,t)+c(\boldsymbol{x},t),
    \\
    U_i(\boldsymbol{x},t)
    &=
    \overline{U}_i(z,t)+u_i(\boldsymbol{x},t).
\end{align}
\end{subequations}
Here, $\overline{C}(z,t)$ and $\overline{U}_i(z,t)$ denote averages over the horizontal $(x,y)$ plane, whereas $c(\boldsymbol{x},t)$ and $u_i(\boldsymbol{x},t)$ denote the corresponding fluctuations. For the present configuration, the mean velocity vanishes, such that $\overline{U}_i(z,t)=0$.

At this stage, it is useful to introduce the width of the RT mixing zone, denoted by $L$, which is the primary quantity to be reproduced from the dataset. Following \citet{Andrews_1990}, this width is defined in terms of the mean concentration profile as
\begin{equation}
    L(t)
    =
    6\int_{-\infty}^{\infty}
    \overline{C}(z,t)\bigl[1-\overline{C}(z,t)\bigr]
    \,\mathrm{d}z.
    \label{eq:mixing_length_def}
\end{equation}
The prefactor $6$ ensures that $L$ coincides with the geometrical thickness of a piecewise-linear mean concentration profile, as is commonly observed in turbulent mixing layers.

Accordingly, the profiles of turbulent kinetic energy and its dissipation rate are defined as
\begin{subequations}
\begin{align}
    k(z,t)
    &=
    \frac{1}{2}
    \overline{u_i u_i},
    \\
    \varepsilon(z,t)
    &=
    \nu
    \overline{\partial_j u_i\,\partial_j u_i}.
    \label{eq:K_Eps_profile_definition}
\end{align}
\end{subequations}

In addition to the mean concentration, turbulent kinetic-energy, and dissipation profiles, which are the main quantities of interest in this study, we introduce the vertical component of the turbulent kinetic energy, defined by
\begin{equation}
    k_{\parallel}(z,t)
    =
    \frac{1}{2}
    \overline{u_z^2}.
    \label{eq:vertical_tke}
\end{equation}
The vertical turbulent mass flux and concentration variance are defined by
\begin{align}
    \mathcal{F}(z,t)
    &=
    \overline{u_z c},
    \\
    b(z,t)
    &=
    \overline{c^2}.
    \label{eq:F_c2_profile_definition}
\end{align}
These additional quantities are used to enforce the physical consistency of the reduced-order models.

\subsection{Spatially integrated quantities (0D)}

In addition to the one-dimensional profiles, spatially integrated quantities are considered throughout this work. These zero-dimensional observables provide a compact description of the global evolution of the instability. For a generic profile $q(z,t)$, we define
\begin{equation}
    Q(t)
    =
    \frac{1}{L(t)}
    \int_{-\infty}^{\infty}
    q(z,t)\,\mathrm{d}z.
    \label{eq:zero_dimensional_quantity}
\end{equation}
The zero-dimensional turbulent kinetic energy, dissipation rate, and mass flux are denoted by $K$, $E$, and $F$, respectively.

\subsection{Phenomenology of the RT dynamics}

The evolution of a Rayleigh--Taylor mixing layer proceeds through several distinct regimes: the linear regime, the weakly nonlinear regime, the transition to turbulence, and, at late times, the self-similar regime \citep{Sharp1984}. These regimes have been extensively studied; see, in particular, \citet{Thevenin_2025_Physica}, who proposed a simple phenomenological model describing their evolution. In this section, we first present the evolution of the zero and one-dimensional quantities and then discuss the late-time scaling laws that underpin the analysis developed throughout this manuscript.

\subsubsection{Example of trajectory 0D/1D dynamics}

\begin{figure}[H]
    \centering
    \includegraphics[width=1.\textwidth]
    {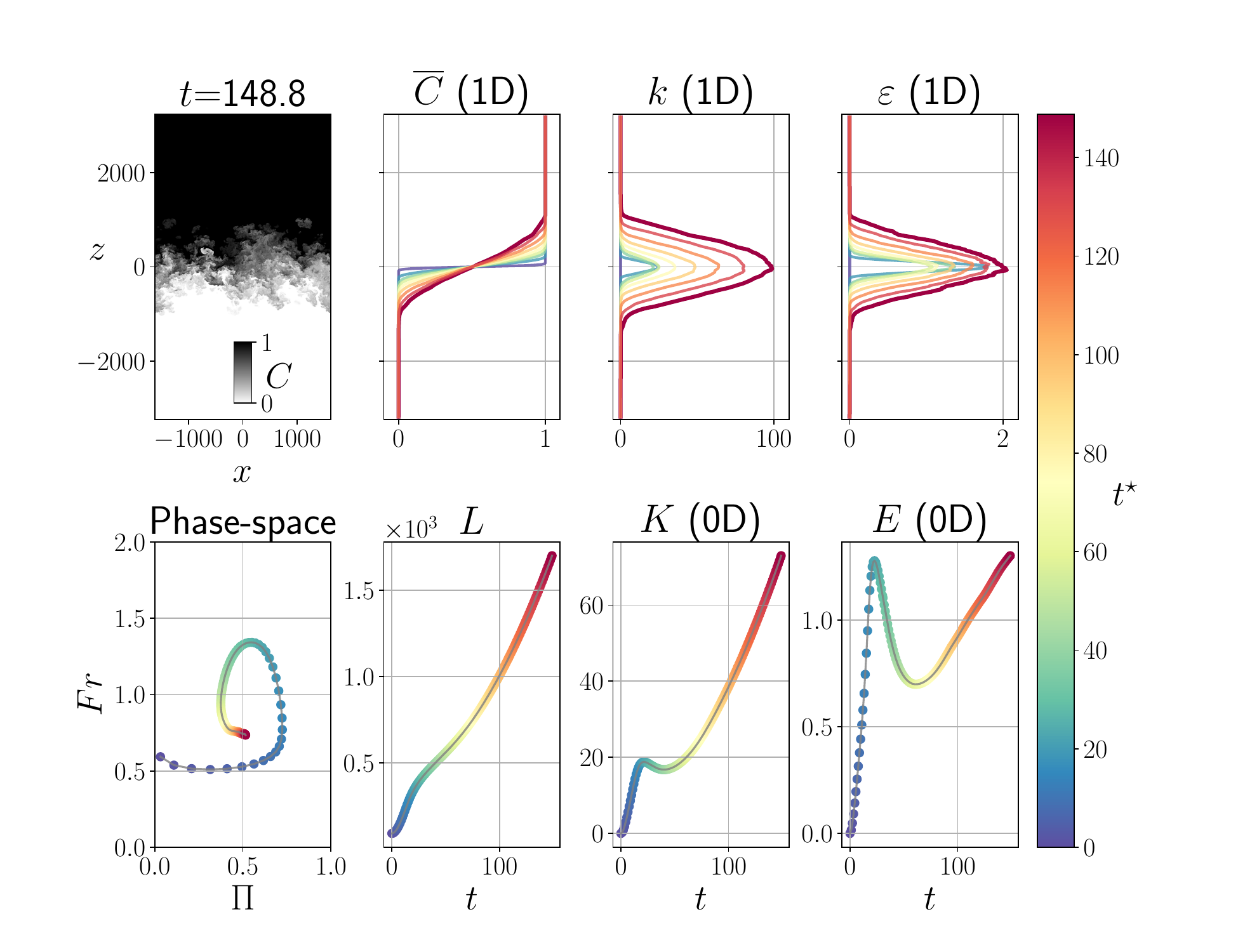}
    \caption{Evolution of a representative Rayleigh--Taylor trajectory with initial-condition parameters $\mathsf{R}=25.2$, $\mathsf{B}=1.2$, $\mathsf{S}=3.0$, and $\mathsf{D}=3.0$, from the initial development of the mixing layer to the turbulent regime. The upper-left panel shows a two-dimensional vertical slice of the instantaneous concentration field at $t=148.8$, while the remaining upper panels show the temporal evolution of the one-dimensional mean-concentration, turbulent kinetic-energy, and dissipation profiles. The lower-left panel represents the same trajectory in the $(\Pi,Fr)$ phase space, with $\Pi$ and $Fr$ defined in Eqs.~\eqref{eq:energy_ratio} and~\eqref{eq:turbulent_froude_number}, respectively. The remaining lower panels show the temporal evolution of the mixing length $L$, zero-dimensional turbulent kinetic energy $K$, and dissipation $E$. Colors indicate the dimensionless time $t$.}
    \label{fig:2D_slices_with_profiles}
\end{figure}

Figure~\ref{fig:2D_slices_with_profiles} illustrates the temporal evolution of the one-dimensional profiles of the mean concentration, turbulent kinetic energy and its dissipation rate, together with their zero-dimensional counterparts. These quantities form the basis of most Reynolds-averaged Navier--Stokes (RANS) turbulence models.

As the initial multimode perturbation grows, dominant spikes and bubbles develop and subsequently undergo secondary instabilities, ultimately triggering the transition to turbulence. This process is marked by a peak in the dissipation rate, also referred to as an enstrophy blow-up, which reflects a transient imbalance in the transfer of energy across scales. Notably, this phenomenology is highly sensitive to the initial conditions, which are characterized here by the parameters $\mathsf{R}$, $\mathsf{B}$, $\mathsf{S}$, and $\mathsf{D}$.

Capturing the dependence of these trajectories on the initial conditions is, therefore, one of the primary objectives of the surrogate model. At late times, the Rayleigh--Taylor mixing layer enters a self-similar regime, whose characteristic scaling laws are examined in greater detail below.

\subsubsection{Self-similar scaling}
\label{sec:self_similarity}

The late-time self-similar regime of Rayleigh--Taylor mixing, which can be anticipated from dimensional analysis \citep{Youngs_1984}, is characterized by a quadratic growth of the mixing-layer width: 
\begin{equation} L(t) \underset{t \to +\infty}{\sim} 2\alpha_\infty \left(t-t_\infty\right)^2, \label{eq:self_similar_mixing_width} 
\end{equation} 
where $\alpha_\infty \approx 0.02$ is the asymptotic self-similar growth coefficient, which has been shown to exhibit universal behavior in our database \citep{Thevenin_2025_Physica}. By contrast, $t_\infty$ denotes the virtual origin of the self-similar solution. Because it is closely related to the transition to turbulence, $t_\infty$ remains strongly dependent on the initial conditions.
It follows directly that the growth velocity of the mixing layer scales linearly with the shifted time: \begin{equation} \dot{L}(t) \underset{t \to +\infty}{\sim} 4\alpha_\infty \left(t-t_\infty\right). \label{eq:self_similar_mixing_velocity} \end{equation}

Introducing the self-similar coordinate \begin{equation} \xi = \frac{z}{L(t)}, \label{eq:xi} \end{equation} and appropriately rescaling the various one-dimensional profiles, the resulting quantities are expected to become independent of time and to collapse, at sufficiently late times, onto corresponding self-similar profiles \(f(\xi)\) (see also \cite{Ristorcelli_Clark_2004}). More precisely, 

\begin{subequations} \label{eq:self_similar_quantities} \begin{align} \widetilde{k}(\xi,t) &= \frac{k\!\left(\xi L(t),t\right)}{L(t)} \xrightarrow[t\to\infty]{} f_k(\xi), \label{eq:self_similar_quantities_K} \\ \widetilde{\varepsilon}(\xi,t) &= \frac{\varepsilon\!\left(\xi L(t),t\right)} {\sqrt{L(t)}} \xrightarrow[t\to\infty]{} f_{\varepsilon}(\xi), \label{eq:self_similar_quantities_eps} \\ \widetilde{\mathcal{F}}(\xi,t) &= \frac{\mathcal{F}\!\left(\xi L(t),t\right)} {\sqrt{L(t)}} \xrightarrow[t\to\infty]{} f_{\mathcal{F}}(\xi), \label{eq:self_similar_quantities_F} \\ \widetilde{C}(\xi,t) &= \overline{C}\!\left(\xi L(t),t\right) \xrightarrow[t\to\infty]{} f_C(\xi), \label{eq:self_similar_quantities_C} \\ \widetilde{b}(\xi,t) &= b\!\left(\xi L(t),t\right) \xrightarrow[t\to\infty]{} f_b(\xi). \label{eq:self_similar_quantities_c2} \end{align} \end{subequations}

\begin{figure}[H]
    \centering
    \includegraphics[width=0.8\textwidth]
    {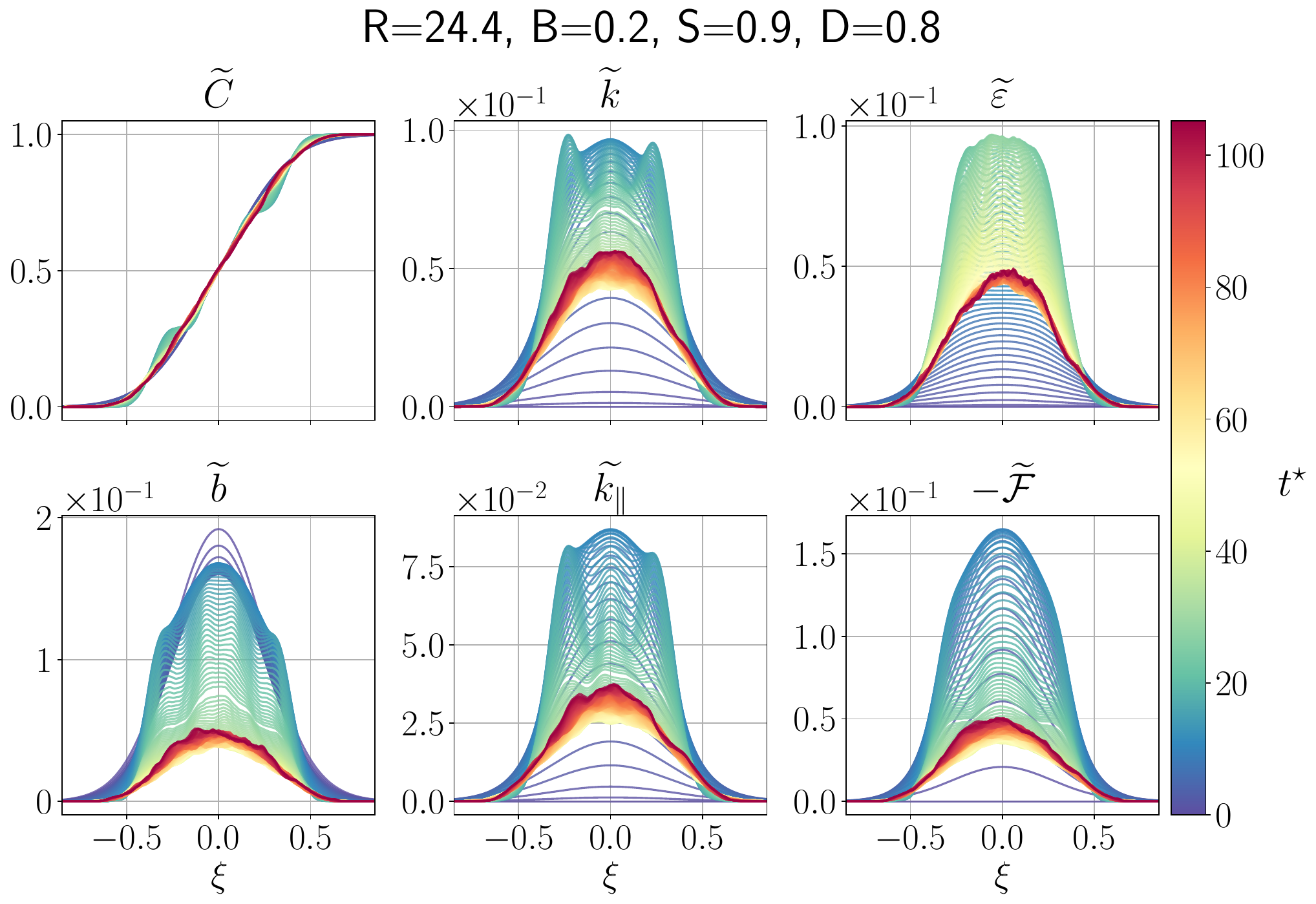}
    \caption{
        Evolution of the self-similar profiles for a representative
        trajectory. The late-time profiles, shown in red, approach their
        asymptotic self-similar shapes.
    }
    \label{fig:self_similar_profiles_all_Q}
\end{figure}

In Fig.~\ref{fig:self_similar_profiles_all_Q}, we show the collapse of the one-dimensional self-similar profiles obtained from a similar trajectory previously depicted in Fig.~\ref{fig:2D_slices_with_profiles}. As expected, the mean concentration profile, $\widetilde{C}$, approaches a nearly piecewise-linear shape, whereas the other turbulent quantities converge toward approximately parabolic profiles.

During the transition to turbulence, near the peak of dissipation, more complex profile shapes may emerge. Localized vortical structures generated by secondary instabilities during the breakup of bubbles and spikes can produce bumps or flattened regions in the one-dimensional profiles near the centerline of the mixing zone. These structural variations are also intended to be captured by the surrogate models.

%
%
\section{POD-based surrogate modeling framework}\label{sec2}

This first strategy places nonlinearity in the temporal dynamics rather than in the profile representation. The one-dimensional fields, expressed using the self-similar scaling defined in Eq.~\eqref{eq:self_similar_quantities}, are projected onto a truncated Proper Orthogonal Decomposition (POD) basis. This constitutes a linear encoding--decoding step: both the reduction of the profiles to modal amplitudes and their reconstruction are linear operations. The evolution of these amplitudes as functions of time and the initial-condition parameters is then learned by a physics-informed neural network (PINN), which provides an explicitly nonlinear model of the reduced dynamics.
%
%
%
%
\subsection{Physics-regulated POD}
%
%
%
%
%
We employ the POD to construct a reduced representation of the self-similar profiles defined in Eq.~\eqref{eq:self_similar_quantities}. A generic transformed quantity is represented in an $r$-dimensional modal space as
\begin{equation}
    \widetilde{Q}(\xi, t)
    \approx
    \sum_{i=1}^{r}
    \phi_i^{\widetilde{Q}}(\xi) a_i^{\widetilde{Q}}(t),
    \label{eq:POD_decomp_definition}
\end{equation}
where $\phi_i^{\widetilde{Q}}$ are the spatial modes and $a_i^{\widetilde{Q}}$ their associated temporal amplitudes. Because the dominant temporal scaling has already been removed, $\widetilde{Q}$ approaches the stationary profile $f_Q(\xi)$ in the self-similar regime and the corresponding modal amplitudes remain bounded at late times.
The POD identifies the set of modes $\{\phi_i^{\widetilde{Q}}\}_{i=1}^r$ that provides the optimal orthogonal basis for a quantity with the norm induced by the inner product
\begin{equation}
	\langle f, g \rangle = \int_{\Omega} f(\xi) \cdot g(\xi)\, d\xi,
	\label{eq:inner_product}
\end{equation}
which is defined on the spatial domain $\Omega=[-0.85, 0.85]$ for profiles $f$ and $g$.
We approximate this integral using a Simpson quadrature, yielding a weighted sum on the discrete approximations of $f$ and $g$ using 201 points.

The POD modes are computed using the method of snapshots combined with singular value decomposition (SVD) \citep{Sirovich_1987}. This decomposition provides the optimal rank-$r$ approximation of the snapshot matrix in the $L_2$ sense \citep{Eckart_1936}.

The objective is to identify modes that efficiently represent the complete ensemble of initial conditions considered in the database, rather than a single trajectory. Because the different quantities exhibit distinct spatial structures and symmetries, a common basis would provide no clear reduction advantage and would compromise the optimal rank-$r$ approximation obtained for each field separately. We therefore construct an independent POD basis for each quantity, thereby improving reconstruction accuracy for a fixed number of retained modes.

To this end, a snapshot matrix is constructed independently for each rescaled quantity expressed self-similarly from 10,000 profiles uniformly sampled over the trajectories and time steps of the training set. These matrices represent approximately 10\% of the available profiles for each quantity.

%
%
%
\paragraph{Self-similar representation}\mbox{}
%

The POD is performed on the rescaled quantities $\widetilde{C}$, $\widetilde{k}$, $\widetilde{k}_{\parallel}$, $-\widetilde{\mathcal{F}}$, $\widetilde{\varepsilon}$, and $\widetilde{b}$. The transformations defined in Eq.~\eqref{eq:self_similar_quantities} map the expanding physical profiles onto the fixed coordinate $\xi$ and remove their known growth with the mixing length; $\widetilde{k}_{\parallel}$ is scaled in the same manner as $\widetilde{k}$.

This formulation prevents late-time snapshots from dominating the POD solely because of their larger physical magnitude. More importantly, the transformed profiles become time independent in the asymptotic self-similar regime. The POD modes can therefore separate the stationary late-time structure from the transient deformations associated with the instability development and transition to turbulence.

%
%
%
%
%
%
%
\paragraph{Symmetry}\mbox{}
%

Under the weak-density-contrast approximation, the Rayleigh--Taylor instability is statistically symmetric about $\xi=0$. Accordingly, $\widetilde{k}$, $\widetilde{k}_{\parallel}$, $-\widetilde{\mathcal{F}}$, $\widetilde{\varepsilon}$, and $\widetilde{b}$ are treated as even quantities, whereas $\widetilde{C}$ is treated as odd. Because individual DNS realizations do not satisfy these symmetries exactly, each snapshot is replaced by its appropriate symmetric component before applying the POD.

This preprocessing reduces the effective number of degrees of freedom and guarantees that every profile reconstructed from the POD basis satisfies the required symmetry, thereby preventing spurious asymmetric predictions and preserving physical consistency.
%
%
\paragraph{Realizability}\mbox{}\label{section:realizability}
%
%

Some of the self-similar quantities are positive because they are defined from squared fluctuations ($\widetilde{k}$, $\widetilde{k}_{\parallel}$, $\widetilde{\varepsilon}$, and $\widetilde{b}$). Their low-rank reconstruction should therefore preserve realizability, although a standard POD expansion provides no such guarantee.

To retain a linear SVD-based encoding while enforcing non-negativity after reconstruction, the POD is applied to the square root of these quantities. The resulting representations are
\begin{equation}
\widetilde{Q}(\xi,t) \approx
\begin{cases}
\left(\displaystyle\sum_{i=1}^{r} \phi_i^{\widetilde{Q}}(\xi)\, a_i^{\widetilde{Q}}(t)\right)^2,
& \text{if } \widetilde{Q} \in \left\{\widetilde{k},\widetilde{k}_{\parallel},\widetilde{\varepsilon}, \widetilde{b}\right\}, \\[1.2em]

\displaystyle\sum_{i=1}^{r} \phi_i^{\widetilde{Q}}(\xi)\, a_i^{\widetilde{Q}}(t),
& \text{if } \widetilde{Q}\in \left\{ \widetilde{C},-\widetilde{\mathcal{F}}\right\}.
\end{cases}
\label{eq:POD_decomp}
\end{equation}

The square-root transformation is applied directly to the sampled self-similar snapshots before the SVD. One POD basis is thus obtained for each of $\sqrt{\widetilde{k}}$, $\sqrt{\widetilde{k}_{\parallel}}$, $\sqrt{\widetilde{\varepsilon}}$, $\sqrt{\widetilde{b}}$, $\widetilde{C}$, and $-\widetilde{\mathcal{F}}$.\\

No temporal or ensemble mean profile is subtracted before the decomposition. In self-similar variables, the late-time state is stationary but nonzero. Removing the mean would therefore eliminate a substantial part of the asymptotic profile that the ROM is intended to reproduce. Retaining it allows the leading POD mode to represent the stationary self-similar structure, while the higher-order modes describe transient departures from that state.
%
%
%
%
%
%
%
%
%
%
%
%
\subsection{Identification of self-similar and transitional modes}
%
%
%
%
%
%
%
The POD framework introduced in the previous section forms the basis of the reduced-order model employed to predict the profile evolution as a function of the initial conditions. To determine an appropriate truncation rank, a modal analysis of the training database is first performed.

The POD is computed from randomly selected self-similar snapshots of the training set. For each quantity, the amplitude associated with the leading mode $\phi_1^{\widetilde{Q}}$ approaches a nonzero asymptotic value once the self-similar regime is reached, while the amplitudes of the higher-order modes decay. Consequently, the leading mode provides a direct modal representation of the stationary self-similar profile: up to a multiplicative factor given by its asymptotic amplitude, $\phi_1^{\widetilde{Q}}$ captures the asymptotic profile shape (see Appendix~\ref{app:interpret_POD_modes}).

All POD modes are presented in Appendix~\ref{app:POD_details}, Fig.~\ref{fig:modes_comparaison_1_a_5_symm}, while the temporal amplitudes of the first two modes for all quantities are shown in Fig.~\ref{fig:amp_pod_mode_1_2}.

The temporal amplitudes associated with the leading modes also provide a direct representation of the corresponding zero-dimensional quantities obtained by spatial averaging of the profiles. Consistently with the usual interpretation of POD decompositions, the first mode captures the dominant average structure of the dataset. Indeed, its temporal amplitude times the mixing layer width $L(t)a^{\widetilde Q}_1(t)$ closely follows the evolution of the corresponding spatially averaged $0 \mathcal{D}$ quantity (see Appendix \ref{app:0D_Q_POD}), with relative errors around $1\%$ over the entire dataset (about $0.1\%$ mean error using all modes).

The secondary modes, $\phi^{\widetilde Q}_{2\leq i\leq5}$, display more complex spatial structures, characterized by localized extrema near the boundaries of the mixing layer. Their temporal amplitudes indicate that they are predominantly active during the early stages of the instability ($t<60$), while they become negligible compared with the leading mode at later times. Moreover, their amplitudes tend to zero in the turbulent regime, suggesting that they play only a minor role in the asymptotic dynamics. A plausible physical interpretation is that these modes capture the initialization phase and the transition towards fully developed turbulence.

The higher-order modes, $\phi^{\widetilde Q}_{i>5}$, exhibit increasingly oscillatory structures that resemble Fourier modes. Their amplitudes remain small throughout the simulations and do not exhibit any clear dependence on the initial conditions $(\mathsf{R},\mathsf{B},\mathsf{S},\mathsf{D})$. Furthermore, they do not admit an obvious physical interpretation and primarily capture high-frequency fluctuations associated with the stochastic nature of the trajectories, such as phase differences between the initially imposed perturbation modes.\\

Based on this modal analysis, five modes are retained for $\widetilde{k}$, $\widetilde{k}_{\parallel}$, $\widetilde{\varepsilon}$, $\widetilde{b}$, and $-\widetilde{\mathcal{F}}$, whereas four modes are sufficient for the mean concentration $\tilde{C}$. This selection provides a compromise between compactness, physical interpretability, and reconstruction accuracy. The resulting truncation error is evaluated on the test set in Appendix~\ref{app:reconstruct_pod_vs_pinn}, Fig.~\ref{fig:relat_err_pod_time}, where it is compared with the complete POD--PINN error.
%
%
%
%
%
%
%
\paragraph{Summary}\mbox{}\label{sec:summary_pod}

\begin{figure}[H]
    \centering
    \includegraphics[width=1.0\textwidth]{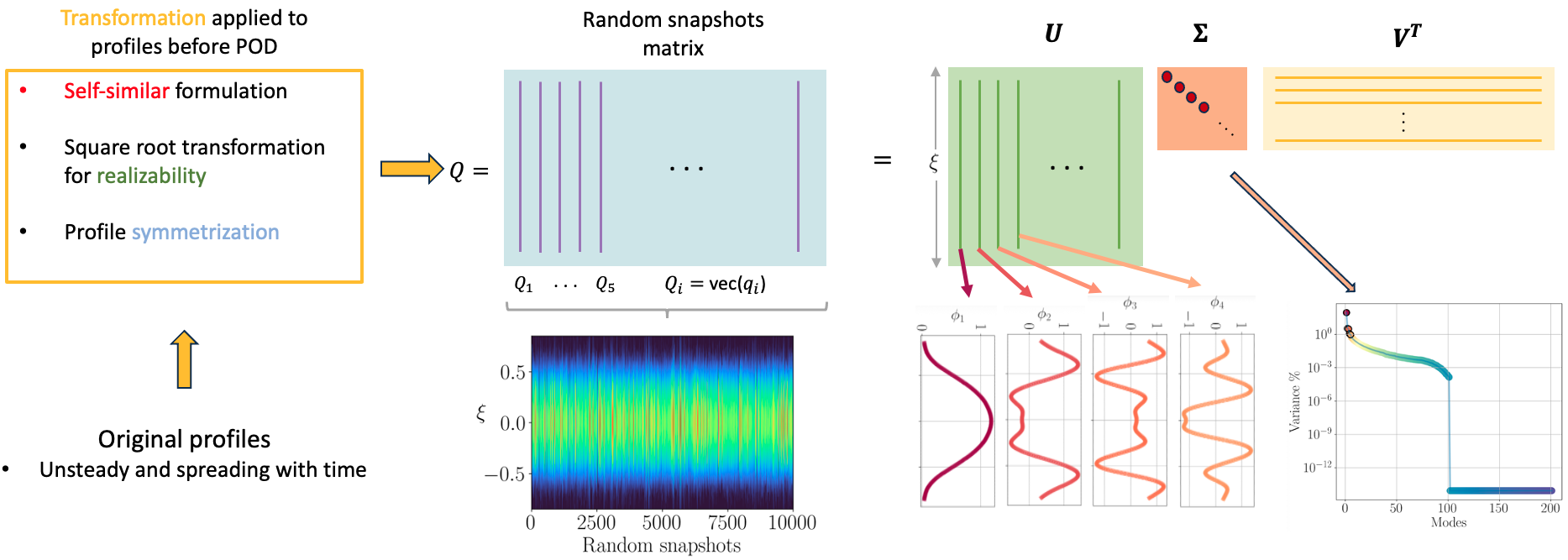}
    \caption{\label{fig:pipeline_POD}{Construction of the POD basis for the self-similar turbulent kinetic energy $\widetilde{k}$. Each physical snapshot is mapped onto the fixed self-similar coordinate $\xi$ and normalized by the corresponding mixing length $L$. A pointwise square-root transformation is then applied before the SVD to ensure a non-negative reconstruction after squaring. In the self-similar regime, the transformed profiles become stationary and remain bounded.}}
\end{figure}

To summarize, the construction of the POD modes for each quantity proceeds as follows (also illustrated in Fig.~\ref{fig:pipeline_POD}):
\begin{enumerate}
\item Randomly draw 10,000 profiles of a given quantity, uniformly sampled across trajectories and time steps, to form the snapshot matrix.
\item Express all snapshots in their self-similar form: $\widetilde{C}$, $\widetilde{k}$, $\widetilde{k}_{\parallel}$, $-\widetilde{\mathcal{F}}$, $\widetilde{\varepsilon}$, and $\widetilde{b}$; see Eq.~\eqref{eq:self_similar_quantities}.
\item Apply a square-root transformation to $\widetilde{k}$, $\widetilde{k}_{\parallel}$, $\widetilde{\varepsilon}$, and $\widetilde{b}$ to ensure non-negative reconstructions.
\item Enforce the statistical symmetry of the profiles: retain the even component of the turbulent quantities and the odd component of $\widetilde{C}$.
\item Perform the singular value decomposition of the snapshot matrix. The POD modes $\phi_i^{\widetilde{Q}}$ are the columns of the matrix $\mathbf{U}$.
\item Six sets of POD modes are obtained, one for each quantity.
\item Retain 5 modes for $\widetilde{k}$, $\widetilde{k}_{\parallel}$, $-\widetilde{\mathcal{F}}$, $\widetilde{\varepsilon}$, and $\widetilde{b}$, and 4 modes for $\widetilde{C}$.
\end{enumerate}
%
%
%
\subsection{Physics-informed surrogate model in reduced space}
%
%
To exploit the reduced-order representation defined in Eq.~\eqref{eq:POD_decomp} and generate profile predictions from arbitrary initial conditions, a surrogate model is required to map the initial parameters $(\mathsf{R},\mathsf{B},\mathsf{S},\mathsf{D})$ to the corresponding temporal POD amplitudes.

This mapping is highly nonlinear, in contrast with the linear encoding-decoding step relating the profiles to their POD amplitudes. Constructing an explicit analytical relation between the initial conditions and the POD coefficients remains challenging.

Moreover, the problem is intrinsically non-unique. The four non-dimensional parameters provide a statistical characterization of the initial perturbation but do not contain information about the phase distribution of the individual perturbation modes. Therefore, different initial perturbations characterized by identical values of $(\mathsf{R},\mathsf{B},\mathsf{S},\mathsf{D})$ may, in principle, lead to different instantaneous evolutions. The surrogate model must therefore learn the dominant statistical trends of the dynamics rather than reproduce the exact trajectory associated with a fully specified initial condition.

To address this challenge, a physics-informed neural network (PINN) framework \citep{Raissi_2019} is introduced to learn the evolution of the POD amplitudes from the initial conditions. By combining data-driven learning with physical constraints, the proposed approach provides a complete parametric reduced-order model. Rather than explicitly integrating a truncated Galerkin system, the reduced dynamics are directly approximated by the neural network while remaining constrained by the governing equations projected onto the POD basis. This strategy naturally accommodates variations of the initial-condition parameters while largely circumventing the closure problem inherent to projection-based ROMs, since the unresolved modal interactions are learned implicitly rather than modeled explicitly. As a result, it also alleviates many of the stability issues commonly encountered in truncated Galerkin systems, while preserving the physical structure provided by the reduced governing equations.
%
%
%
\paragraph{Model architecture and preprocessing strategies}\mbox{}

The proposed PINN is designed to predict the temporal evolution of the reduced-order variables from the initial conditions at a given time. The network inputs consist of the four non-dimensional parameters characterizing the initial perturbation and time:
\begin{equation}
    \mathbf{x} =
    (\mathsf{R}, \mathsf{B}, \mathsf{S}, \mathsf{D}, t)^T .
\end{equation}
It outputs the mixing-layer width and the POD amplitudes associated with each quantity of interest:
\begin{equation}
    \mathbf{y} =
    \left(
    L,
    \{a_i^{\widetilde{k}}\}_{i=1}^{5},
    \{a_i^{\widetilde{k}_{\parallel}}\}_{i=1}^{5},
    \{a_i^{\widetilde{\varepsilon}}\}_{i=1}^{5},
    \{a_i^{\widetilde{\mathcal{F}}}\}_{i=1}^{5},
    \{a_i^{\widetilde{b}}\}_{i=1}^{5},
    \{a_i^{\widetilde{C}}\}_{i=1}^{4}
    \right)^T .
\end{equation}
Therefore, $\mathbf{x}\in\mathbb{R}^{5}$ and $\mathbf{y}\in\mathbb{R}^{30}$. The prediction of the mixing-layer width is essential to reconstruct the physical profiles $Q(z,t)$ rather than their self-similar representation $\widetilde{Q}(\xi,t)$.

Moreover, $L$ cannot be recovered from the concentration field expressed in self-similar coordinates. For this reason, the dimensionless mixing-layer width is explicitly predicted by the network.

The time derivative of the mixing-layer width, $\dot{L}$, is not included as an independent output variable, since it is obtained directly from the predicted $L$ through automatic differentiation.

Several preprocessing strategies and architectural choices are introduced to ease the optimization process. These choices aim at reducing the dynamic range of the variables, exploiting the self-similar properties of the flow, and incorporating physical prior knowledge into the model, see Appendix \ref{app:PINN}.

The PINN architecture is based on a multilayer perceptron composed of three hidden layers with 32 neurons each, resulting in 3294 trainable parameters. 

The training procedure consists of two successive optimization stages. First, the ADAM optimizer, a first-order stochastic gradient-based method, is employed for 200 epochs to obtain a robust initialization of the network parameters. The training is then continued using the Self-Scaled Broyden method, \cite{Urban_2025}. Compared with first-order optimization methods, this quasi-Newton approach exploits information on the local curvature of the loss landscape through an approximate inverse Hessian update. The residual-based attention strategy \cite{Anagnostopoulos_2024}, is also employed to adaptively manage the collocation points, see Appendix \ref{app:colloc_pts}.
%
%
%
%
%
%
%
%
%
%
%
%
%
%
%
%
\paragraph{Construction of the loss function}\mbox{}

In the physics-informed learning framework, the network parameters are optimized by minimizing a composite loss function combining data-driven and physics-informed based contributions:
\begin{equation}
    \mathcal{L} = \mathcal{L}^{Data}+\mathcal{L}^{PI}.
    \label{eq:loss_totale}
\end{equation}
The supervised contribution $\mathcal{L}^{Data}$ measures errors with respect to the DNS quantities projected onto the POD basis, its detailed formulation is given in Appendix~\ref{app:PINN}. Here, we focus on $\mathcal{L}^{PI}$, which imposes physical constraints directly on the predicted mixing length and modal amplitudes. 
The physics-informed contribution comprises three terms:
\begin{equation}
	\mathcal{L}^{PI}
	=
	\mathcal{L}_{\dot{L}>0}
	+
	\mathcal{L}_{k_{0\mathcal{D}}}
	+
	\mathcal{L}_{C_{1\mathcal{D}}},
\end{equation}
These terms respectively constrain the mixing-layer growth, the global turbulent kinetic-energy balance, and the local mean-concentration balance. The latter two introduce physical couplings between several retained amplitudes and therefore complement the pointwise DNS supervision.

The first constraint enforces the monotonic growth of the mixing layer. Since the acceleration considered in the present RTI configuration remains permanently destabilizing, the mixing layer is expected to continuously expand and therefore satisfy $\dot{L}>0$.

The second constraint enforces the zero-dimensional turbulent kinetic-energy balance, whose terms can be expressed directly from the retained POD amplitudes. Because truncation leaves unresolved modal contributions, the associated residual is denoted $r_{k}^{\mathrm{pod}}$ for DNS-projected amplitudes and $r_{k}^{\mathrm{pinn}}$ for PINN predictions.

Rather than introducing an explicit closure model, $r_{k}^{\mathrm{pinn}}$ is evaluated and minimized using only the retained modes. The PINN thereby adjusts their dynamics to compensate for unresolved modal contributions, implicitly learning an effective closure while preserving the global energy balance.

The final physical constraint enforces the one-dimensional advection--diffusion balance, coupling the mean concentration to the turbulent mass flux, expressed in self-similar coordinates through their retained POD representations.

Unlike the previous global constraints, this local relation explicitly involves the POD spatial modes and their derivatives. It therefore promotes consistency between the temporal evolution and the underlying spatial dynamics.

Further details on the loss function formulations are provided in Appendix~\ref{app:PINN}. All technical aspects of the PINN implementation presented in this work are developed within the \textsf{PyTorch} framework \citep{pytorch_2019} using an in-house library.

%
%
%
\section{Autoencoder-based surrogate modeling framework}\label{sec3}

The PINN introduced in Sec.~\ref{sec2} maps the initial-condition parameters and time directly to the quantities of interest. Because the predicted state is not fed back into the model, this approach does not define a temporal propagator and therefore does not provide a framework for analyzing dynamical stability. It is a solution map that maps the set of parameters and a given time to the quantities of interest at that time and that set of parameters. The present framework instead adopts an explicit dynamical-system formulation. The high-dimensional spatiotemporal trajectories are first mapped to a low-dimensional latent space using a nonlinear autoencoder learned directly from the data. Unlike the POD amplitudes of Sec.~\ref{sec2}, the resulting latent coordinates are not assigned an explicit physical interpretation. Their temporal evolution is then constrained to be linear and is governed by a single generator $A$. This construction defines a dynamical system in latent space for which stability analysis is well posed. As with the POD--PINN surrogate, training incorporates the physics-informed constraints detailed in Sec.~\ref{sec3_ae}. Thus, the two frameworks differ primarily in the placement of nonlinearity and in their use of either a static solution map or an explicit dynamical evolution model.
\subsection{Autoencoder with linear latent dynamics}\label{sec3_ae}
The rescaled one-dimensional profiles, Eq.\eqref{eq:self_similar_quantities}, used as inputs, $\mathbf y(\xi,t)\in\mathbb{R}^{n_t\times n_c\times n_\xi}$, stack $n_t$ time steps of $n_c$ channels $\{\widetilde{k},\widetilde{\varepsilon},\widetilde{C},\widetilde{\mathcal F}, \widetilde{k_{\parallel}},\widetilde{b} \}$ on the $n_\xi$-point self-similar grid $\xi$. A transformer-based encoder $\mathcal E$ maps each high-dimensional spatiotemporal trajectory $\mathbf y(\xi,t)$ to a low-dimensional latent space trajectory $\mathbf z(t)\in\mathbb{R}^{n_t\times d}$, with $d\ll n_c n_\xi$, and a decoder $\mathcal D$ of mirrored architecture reconstructs $\widehat{\mathbf y}=\mathcal D(\mathbf z)$. Unlike a standard autoencoder, the latent trajectory is constrained to enforce linear evolution of the trajectory in the low dimensional latent space,
\begin{equation}
\dot{\mathbf z}(t) = A\,\mathbf z(t), \qquad A\in\mathbb{R}^{d\times d},
\label{eq:lasdi_linear_dynamics}
\end{equation}
so that, for every training trajectory, the operator $A$ is obtained by least-square regression of the finite-difference latent derivatives $\dot{\mathbf z}(t_k)$ onto the encoded states $\mathbf z(t_k)$.
The encoder-decoder and this per-trajectory identification of $A$ are trained jointly so that $\mathcal D(\mathbf z)$ reconstructs $\mathbf y$ and $\mathcal E(\mathcal D(\mathbf z))$ is consistent with $\mathbf z$.
As with the POD-PINN surrogate, training is further regularized with soft physical constraints evaluated on the decoded predictions $\widehat{\mathbf y}$: the zero-dimensional kinetic-energy budget Eq.~\eqref{eq:K_0D_eq}, and the mean-concentration advection-diffusion constraint of Eq.~\mbox{\eqref{eq:loss_eq_C0D}}, evaluated on the decoded $\widetilde{C}$ and $\widetilde{\mathcal F}$ channels.

\begin{figure}[H]
    \centering
    \includegraphics[width=0.95\textwidth]{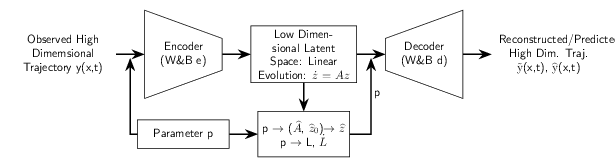}
    \caption{\label{fig:lasdi_schematic}
    Schematic of the LasDI surrogate.
    The encoder maps the observed high-dimensional trajectory $\mathbf y(\xi,t)$ onto a low-dimensional latent space with linear evolution $\dot{\mathbf z}=A\mathbf z$, from which the decoder reconstructs or predicts $\widetilde{\mathbf y}(\xi,t)$, $\widehat{\mathbf y}(\xi,t)$.
    The lower box represents the parametric surrogate of Sec.~\ref{sec3_param}: the physical parameters $\mathbf p$ are mapped to the pair $(\widehat A,\widehat{\mathbf z}_0)$ used to integrate the latent dynamics, and to the mixing length width $L$. The encoder and decoder are further conditioned by the parameter vector $\mathbf p$.}
\end{figure}
\subsection{Parametric prediction of the latent dynamics and mixing length}
\label{sec3_param}
Because $A$ and the initial latent state $\mathbf z_0$ are identified independently for each training trajectory, they can themselves be treated as target quantities of a parametric surrogate, analogous to the role the PINN plays for the POD basis in Sec.~\ref{sec2}. A network $\mathcal P_{\mathbf z}$ maps the physical parameters $\mathbf p=(\mathsf R,\mathsf B,\mathsf S,\mathsf D)$ directly to $(\widehat A(\mathbf p),\widehat{\mathbf z}_0(\mathbf p))$, from which the latent trajectory is obtained by integrating Eq.~\eqref{eq:lasdi_linear_dynamics},
%
%
applied stepwise as $\widehat{\mathbf z}_{k+1}=\exp(\widehat A\,\Delta t)\,\widehat{\mathbf z}_k$. Decoding $\widehat{\mathbf z}(t)$ then yields a full prediction of the profiles for any $\mathbf p$, including parameter combinations absent from the training set, without requiring an encoded reference trajectory.

A second, independent network $\mathcal P_L$ maps the parameters and time to the mixing length and its growth rate, $\mathcal P_L(\mathbf p,t)=(\widehat L,\partial_{t}\widehat L)$, playing the same role as the mixing-length output of the PINN in Sec.~\ref{sec2}.
These predictions are used in implementing the zero-dimensional kinetic-energy budget Eq.~\eqref{eq:K_0D_eq}, and to rescale the self-similar profiles.
In its conditional variant, the encoder and decoder are additionally conditioned by $(\mathbf p,\widehat L,\partial_{t}\widehat L,\Delta t)$ as indicated in Fig.~\ref{fig:lasdi_schematic}. Here, we only present results from the conditional variant since it performs slightly better (2-3\% less relative error).

Given the temporal characteristics of the data, we anticipated being able to fit models with latent space dimension of up to about eight. As such, we considered latent space dimensions of 2, 4, and 8. Results are presented here for latent space dimension of 4; results with latent space dimension of 2 and 8 were only slightly worse.
%
%
\section{Comparison of the reduced-order models}\label{sec:compare_ROM}

This section compares the POD--PINN and LaSDI reduced-order models against the DNS reference data. The analysis considers their predictions of the one-dimensional profiles and associated zero-dimensional quantities, both within and beyond the temporal range covered by the simulations, and evaluates their consistency with the physical constraints considered in this study.

\subsection{Profile dynamics for a representative trajectory}
%
%
%
\begin{figure}[h]
    \centering
    \includegraphics[width=0.95\textwidth]{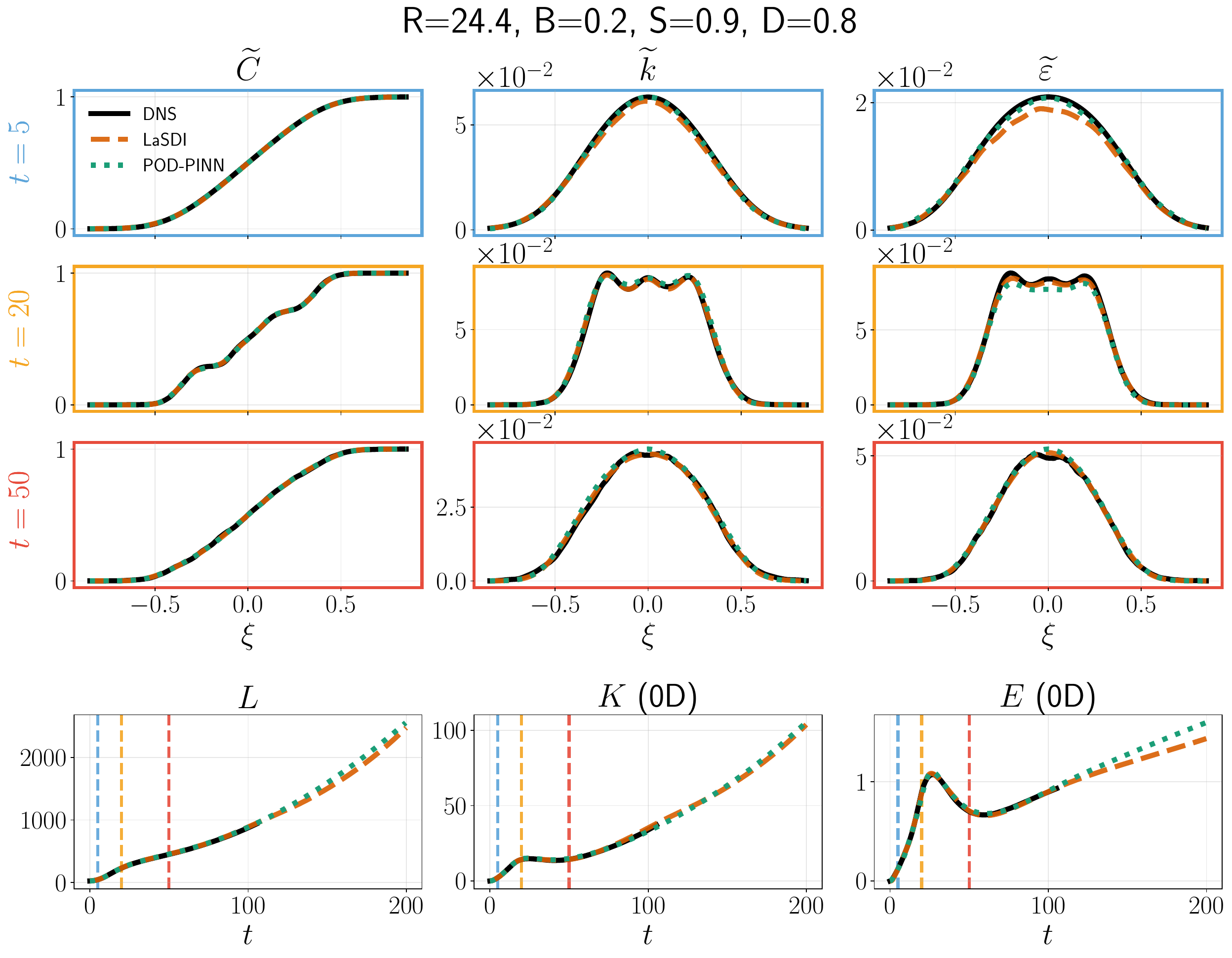}
    \caption{\label{fig:compare_profiles} 
    Comparison of DNS, LaSDI, and POD-PINN predictions for a test simulation. The columns correspond to the mean concentration $\widetilde{C}$, turbulent kinetic energy $\widetilde{k}$, and dissipation rate $\widetilde{\varepsilon}$. The first three rows show instantaneous profiles at selected times $t=5$, $20$, and $50$. The last row displays the evolution of the associated zero-dimensional quantities, including temporal extrapolation beyond the DNS time horizon. The DNS data are available up to $t=105$, whereas both surrogate models are evaluated at later times. The vertical dashed lines and colored frames identify the times selected for the instantaneous profile comparisons.}
\end{figure}

Figure~\ref{fig:compare_profiles} shows that both reduced-order models reproduce the DNS profiles throughout the transition toward turbulence. The mean concentration profiles are nearly indistinguishable from the reference at all three selected times. At $t=20$, LaSDI and POD-PINN also recover the more complex transient structures of the turbulent kinetic energy and dissipation profiles, including the locations of the local extrema and the nonuniform plateaus. The smoother profiles observed at $t=50$ are likewise captured by both models.

The remaining differences primarily concern the profile amplitudes. At $t=5$, LaSDI slightly underestimates the maximum turbulent kinetic energy and, more noticeably, the dissipation peak, whereas POD-PINN remains closer to the DNS. At $t=20$, both models exhibit modest deviations across the turbulent plateaus but preserve their overall structure. At $t=50$, the agreement remains close.

Unlike POD--PINN, whose reconstructions remain confined to the symmetric or antisymmetric subspaces imposed during construction of the POD bases, LaSDI does not enforce reflection symmetry and can therefore represent asymmetric profile components present in individual realizations.

The zero-dimensional quantities confirm the similar accuracy of the two models over the DNS time interval. Beyond the DNS time horizon, the predictions retain the same qualitative trends but gradually separate, with POD-PINN yielding slightly larger late-time values than LaSDI. In particular, the extrapolated integrated dissipation appears to depart from the theoretically expected linear late-time growth. The comparison therefore indicates comparable interpolation performance, while the increasing differences and loss of the expected asymptotic scaling reflect the greater uncertainty associated with temporal extrapolation.

\begin{figure}[h]
    \centering
    \includegraphics[width=1\textwidth]{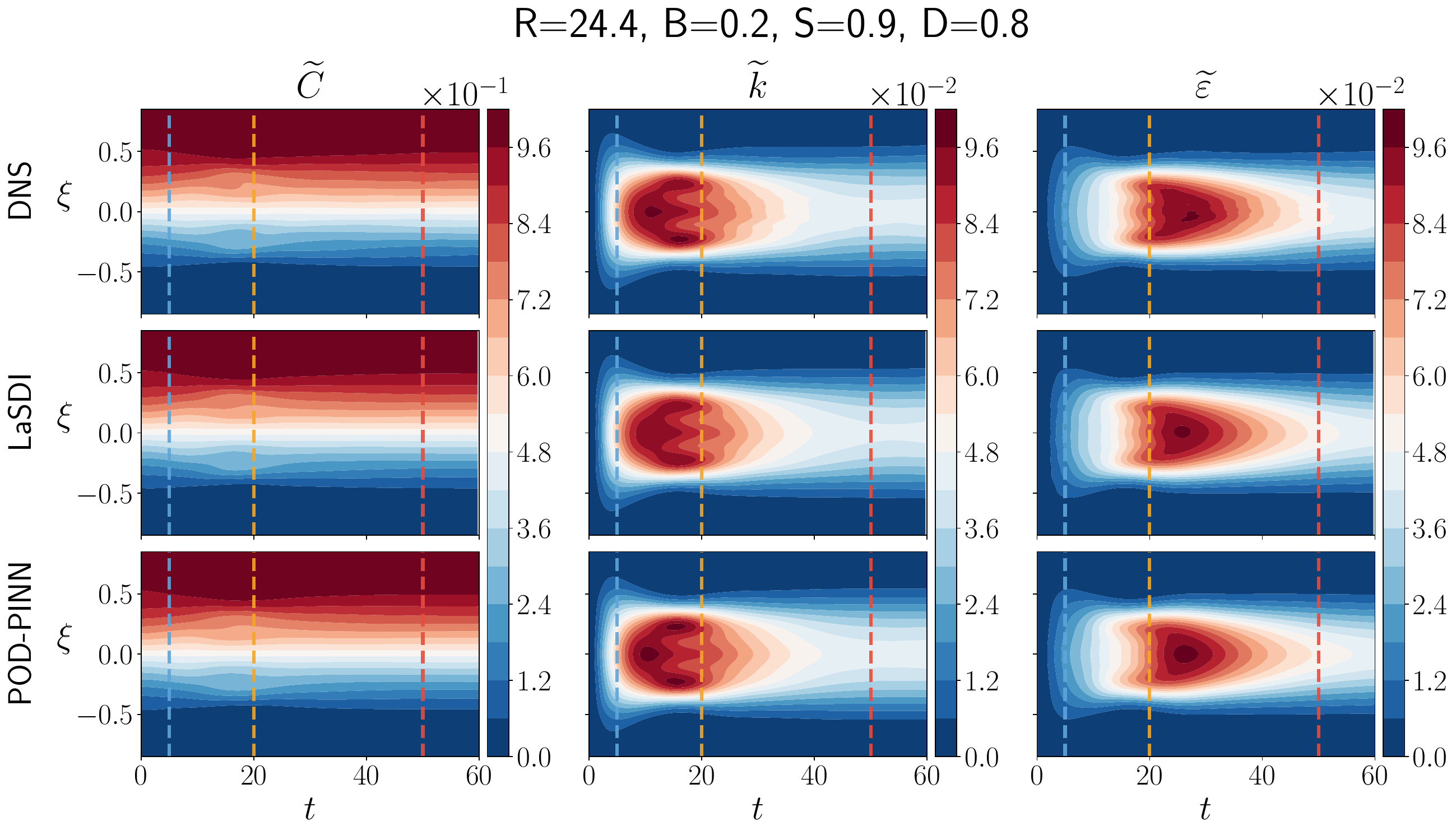}
    \caption{\label{fig:compare_profiles_stack} 
    Comparison of DNS, LaSDI, and POD-PINN predictions for a test simulation. The columns correspond to the self-similar mean concentration $\widetilde{C}$ \eqref{eq:self_similar_quantities_C}, turbulent kinetic energy $\widetilde{k}$ \eqref{eq:self_similar_quantities_K}, and dissipation rate $\widetilde{\varepsilon}$ \eqref{eq:self_similar_quantities_eps}. Each panel represents a space--time matrix obtained by stacking the self-similar one-dimensional profiles at successive times, thereby illustrating the temporal evolution of the fields as a function of the normalized coordinate $\xi$. The rows correspond to the DNS reference solution, the LaSDI surrogate, and the POD-PINN surrogate, respectively. Only the transition-to-turbulence regime is shown. The dashed lines indicate the times selected for the instantaneous profile comparisons in Fig.~\ref{fig:compare_profiles}.}
\end{figure}

The space--time representations in Fig.~\ref{fig:compare_profiles_stack} complement the instantaneous comparisons of Fig.~\ref{fig:compare_profiles} by showing how the profile shapes develop throughout the transition to turbulence. For this trajectory, the DNS fields exhibit pronounced spatial structures during the transient regime, including secondary lobes in the self-similar turbulent kinetic energy and dissipation profiles around $t=20$. Both reduced-order models reproduce the location, persistence, and subsequent attenuation of these structures. The agreement demonstrates that LaSDI and POD-PINN capture not only the global growth of the instability but also the nontrivial spatial dynamics associated with the transition.

Such complex transient profiles are observed in particular for trajectories characterized by $\mathsf{R}\approx25$ and $\mathsf{S}\approx1$. These parameters correspond to an initial interface perturbation dominated by relatively low wavenumbers and having a small initial amplitude, respectively, so that the interface is initially close to a flat configuration. Under these conditions, large coherent structures develop rapidly and accelerate away from the initial interface, accumulating kinetic energy. Their localized transport of kinetic energy and dissipation produces the secondary extrema, or bumps, visible in the one-dimensional profiles during the transition to turbulence regime. As the instability evolves, secondary Kelvin--Helmholtz-type instabilities develop along the shear layers surrounding these structures, promoting the generation of smaller scales and accelerating the transition toward turbulence (around $t=50$). This progressive cascade of spatial scales gradually smooths the profiles and leads to the parabolic self-similar shapes characteristic of the asymptotic regime.

The ability of both ROMs to recover these features is therefore an important result, since they arise from strongly transient and spatially localized dynamics rather than from the smoother late-time self-similar evolution.
%
\subsection{Comparison across initial perturbations}

\begin{figure}[h]
    \centering

    \begin{subfigure}[c]{0.57\linewidth}
        \centering
        \includegraphics[width=\linewidth]{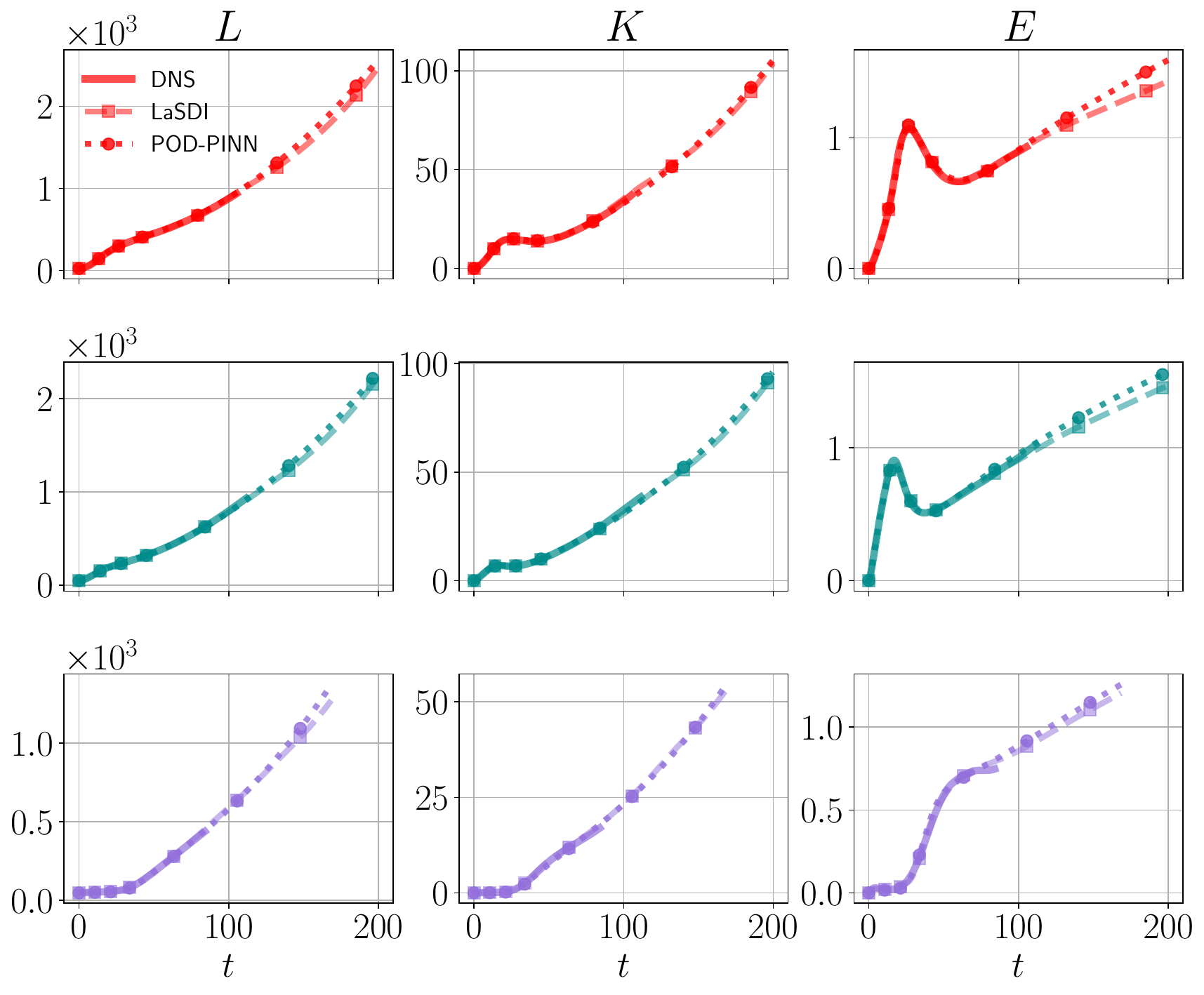}
        \caption{Temporal evolution of the zero-dimensional quantities $K$, $E$, and $L$ obtained from the DNS database and the two surrogate models.}
        \label{fig:0D_comparison}
    \end{subfigure}
    \hfill
    \begin{subfigure}[c]{0.4\linewidth}
        \centering
        \includegraphics[width=\linewidth]{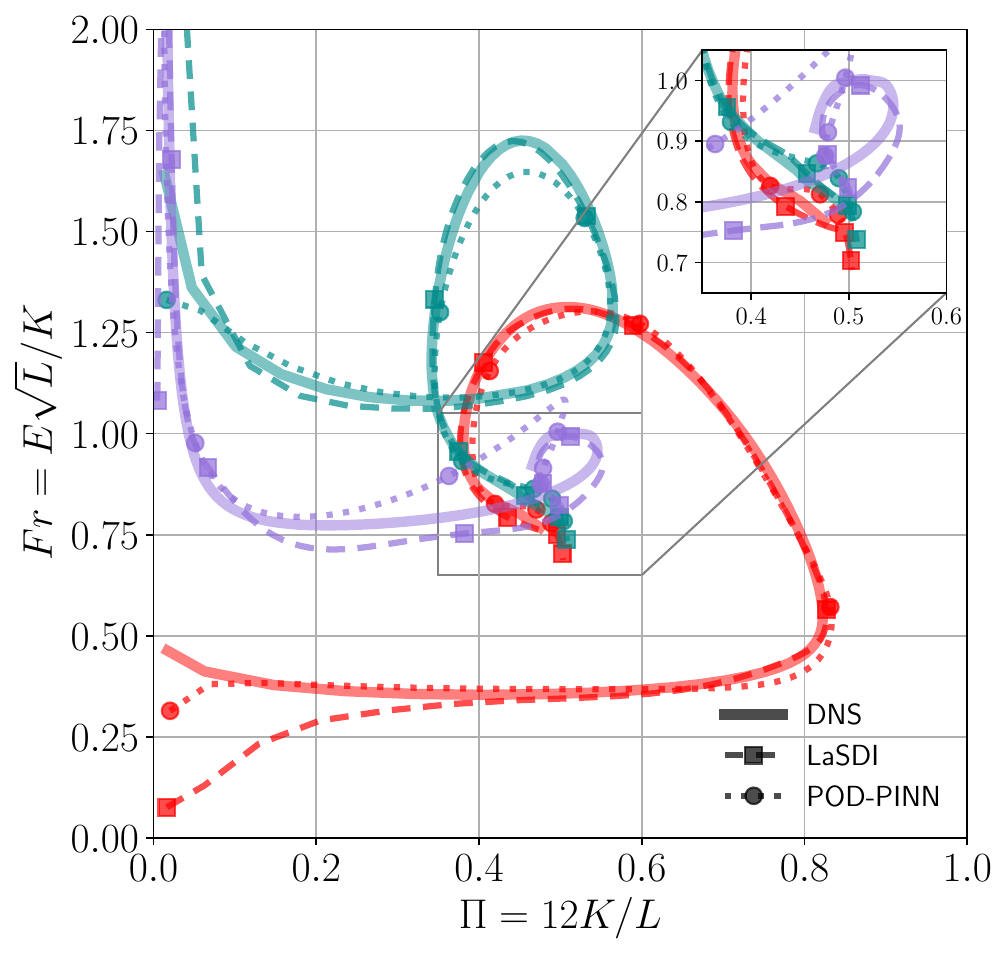}
        \caption{Evolution of the trajectories in the $(\Pi,Fr)$ phase space.}
        \label{fig:Pi_Fr_comparison}
    \end{subfigure}

    \caption{Comparison between three representative trajectories from the test set, comparing the DNS data with the surrogate predictions. The temporal evolution of the main zero-dimensional quantities is shown together with the corresponding trajectories in the $(\Pi,Fr)$ phase space. The red, cyan, and purple curves correspond to inertial, intermediate, and diffusive trajectories, respectively.}
    \label{fig:surrogate_comparison_0D_phase}
\end{figure}

Figure~\ref{fig:surrogate_comparison_0D_phase} compares the zero-dimensional predictions of the two reduced-order models for three representative simulations from the test set. The red trajectory corresponds to the inertial configuration considered in Figs.~\ref{fig:compare_profiles} and~\ref{fig:compare_profiles_stack}, the cyan trajectory represents an intermediate regime, and the purple trajectory corresponds to a diffusion-dominated configuration. 

Additional comparisons between the surrogate models and the DNS are provided in Appendix~\ref{app:additional_rom_comparisons} for trajectories with moderate or low initial Reynolds number $\mathsf R$, whose early-time evolution is primarily diffusion-dominated. This selection illustrates the good performance of the ROMs across markedly different transition dynamics. 

Two dimensionless quantities are used to characterize the trajectories in the phase-space representation. The turbulent Froude number is defined here as
\begin{equation}
    Fr
    =
    \frac{E\sqrt{L}}{K},
    \label{eq:turbulent_froude_number}
\end{equation}
and compares the turbulent frequency, $E/K$, with the characteristic stratification frequency, $1/\sqrt{L}$. The energy ratio
\begin{equation}
    \Pi
    =
    12\frac{K}{L}
    \label{eq:energy_ratio}
\end{equation}
measures the fraction of the available potential energy converted into turbulent kinetic energy. Its derivation from the total-energy balance is detailed in Appendix~\ref{annex:energy_conservation_section}. This quantity is bounded by $0\leq\Pi\leq1$: the upper limit corresponds to a complete conversion of the available potential energy into turbulent kinetic energy in the absence of dissipation, whereas $\Pi=0$ corresponds to the absence of kinetic-energy production from the available potential-energy reservoir.

Figure~\ref{fig:0D_comparison} shows that both models reproduce the main temporal trends of the integrated kinetic energy, dissipation, and mixing length for the three regimes. Their relative performance nevertheless depends on the initial condition and on the predicted quantity.
These results indicate that neither reduced representation is uniformly superior across all dynamical regimes, especially for dissipation, which is particularly sensitive to the rapid development of small scales.

The trajectories in the $(\Pi,Fr)$ phase space, shown in Fig.~\ref{fig:Pi_Fr_comparison}, provide a more compact view of the predicted dynamics. LaSDI has greater difficulty reproducing the initial part of the trajectories, where both dimensionless quantities vary rapidly. For the diffusion-dominated simulation, however, the LaSDI trajectory follows the DNS reference more closely in the phase-space than the POD-PINN prediction. In this phase space, the stable fixed point represents the self-similar state, in which both $\Pi$ and $Fr$ are expected to become constant. The two ROMs preserve this attracting structure. However, the DNS trajectories do not extend sufficiently far into the self-similar regime to identify the reference fixed-point coordinates accurately. The differences between the limiting coordinates predicted by the two models should therefore be interpreted cautiously.
%
%
\subsection{Test-set accuracy and physical consistency}

\begin{figure}[H]
    \centering
    \includegraphics[width=1\textwidth]{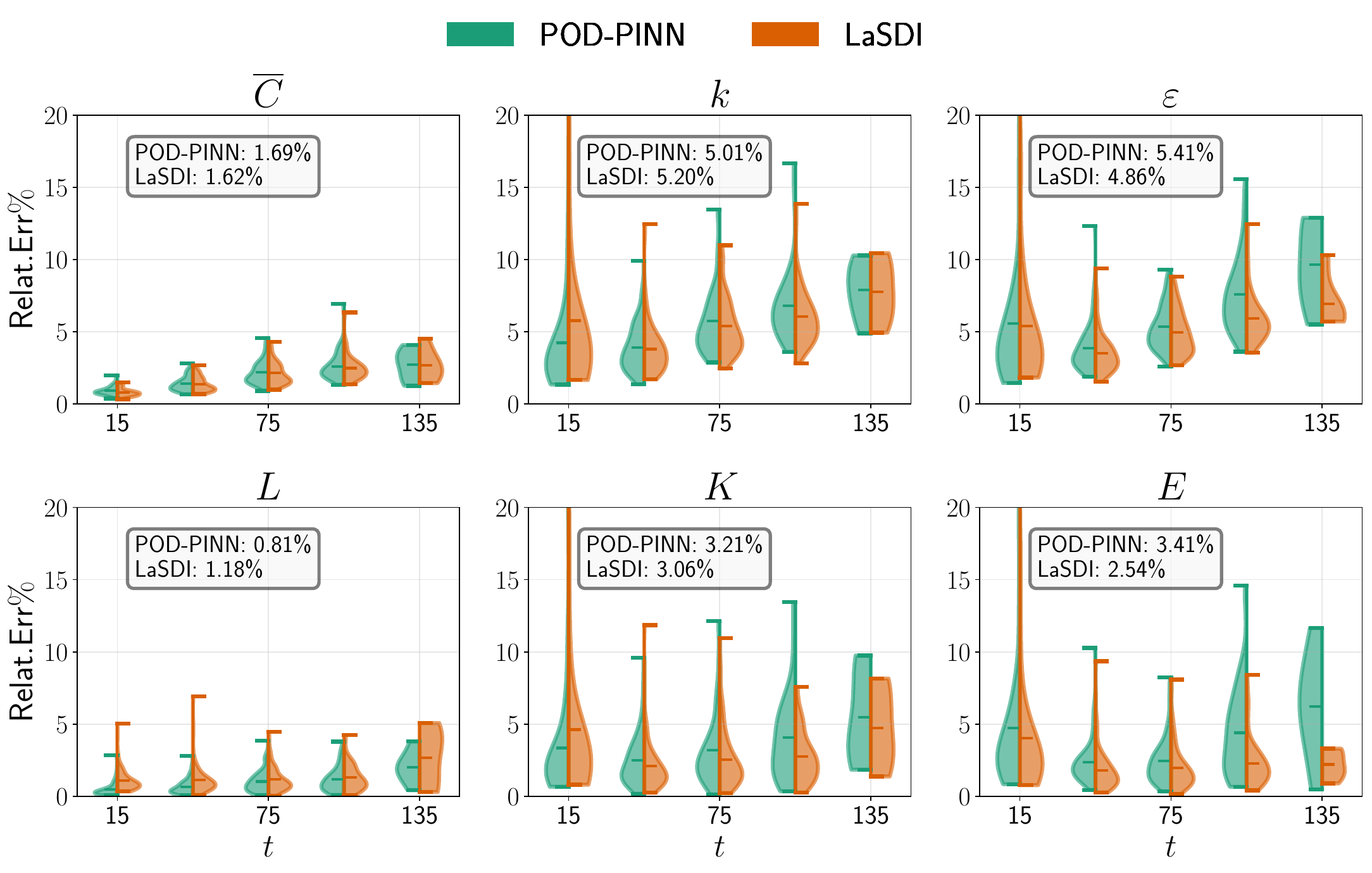}
    \caption{Temporal distributions of the relative errors of the POD--PINN (green) and LaSDI (orange) predictions with respect to the DNS over the test set. Each violin contains one mean error per trajectory and time interval. For the one-dimensional profiles $\overline{C}$, $k$, and $\varepsilon$, the pointwise relative error is first averaged over $\xi \in [-0.5,0.5]$ and then over the snapshots in the interval; restricting the spatial domain avoids divergences near the profile boundaries. For the zero-dimensional quantities $L$, $K$, and $E$, the relative error is evaluated directly from the corresponding scalar time series. The values displayed in each panel are the mean errors over the complete test set.}
    \label{fig:compare_error_time_distrib}
\end{figure}

Figure~\ref{fig:compare_error_time_distrib} compares the two reduced-order models only at times for which DNS reference data are available. The POD-PINN and LaSDI error distributions largely overlap for both the one-dimensional profiles and the associated zero-dimensional quantities, indicating comparable accuracy at the data-supported points. The mean errors are nearly identical for $\overline{C}$, $k$, and $K$, and remain of the same order for the other quantities. The larger errors in the earliest time interval mainly arise from predictions at $t<2$, for which imperfect surrogate initialization is particularly pronounced in diffusion-dominated trajectories.

The modest differences depend on the predicted quantity rather than revealing a systematic advantage of either model. POD-PINN yields a lower mean error for the mixing length $L$, whereas LaSDI is slightly more accurate for the dissipation profile $\varepsilon$ and its integrated counterpart $E$. The distributions generally broaden and shift toward larger errors with time, particularly for the turbulent quantities, but substantial overlap remains between the two models across the temporal intervals.

The latest distributions must nevertheless be interpreted with caution because the DNS trajectories do not all reach the same final time. The number of simulations contributing to each temporal bin therefore decreases with time, making the last intervals less statistically representative and more sensitive to individual trajectories. Moreover, because the errors are evaluated only where DNS data exist, this figure assesses interpolation and reconstruction accuracy rather than long-time extrapolation.

\begin{figure}[H]
    \centering
    \includegraphics[width=0.7\textwidth]{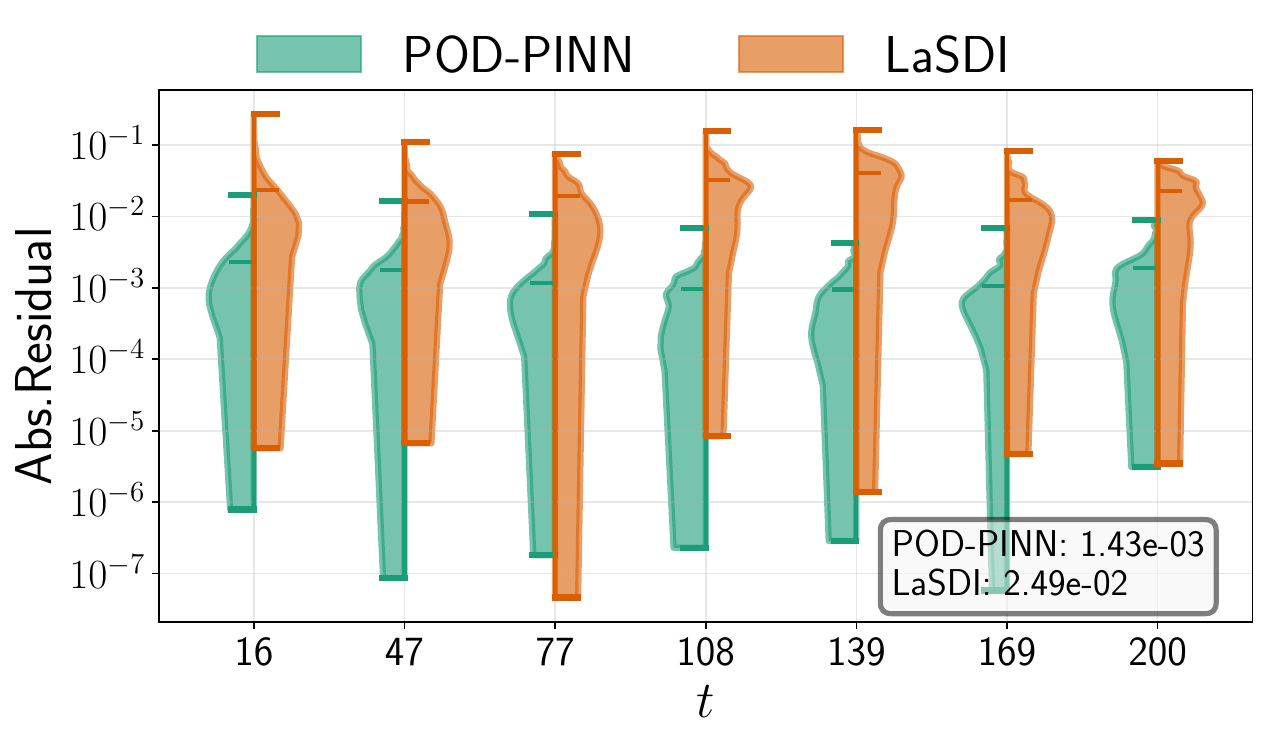}
    \caption{Temporal distributions, on a logarithmic scale, of the absolute residual of the zero-dimensional turbulent kinetic-energy equation~\eqref{eq:K_0D_eq} for the POD--PINN and LaSDI predictions associated with unseen initial conditions $\{\mathsf{R},\mathsf{B},\mathsf{S},\mathsf{D}\}$ from the test set. For both models, the residual is evaluated directly from the predicted quantities. The values displayed in the lower-right corner are the mean residuals over the complete test set and time interval.}
    \label{fig:distrib_err_eq_K_0D}
\end{figure}

The zero-dimensional turbulent kinetic-energy balance~\eqref{eq:K_0D_eq} is imposed as a soft physical constraint during the training of both reduced-order models. Figure~\ref{fig:distrib_err_eq_K_0D} therefore evaluates how well this constraint generalizes to unseen initial conditions and to times extending beyond the DNS data. Because its residual can be computed directly from the surrogate predictions without requiring a DNS reference, it provides a diagnostic of the internal consistency and physical coherence of the extrapolated quantities.

POD--PINN yields systematically smaller residuals than LaSDI over all temporal intervals, including the extrapolation regime. The mean absolute residual is $1.43\times10^{-3}$ for POD--PINN and $2.49\times10^{-2}$ for LaSDI, corresponding to a difference of approximately one order of magnitude. This separation indicates that the POD--PINN predictions remain more accurately coupled through the global kinetic-energy balance, even when evaluated beyond the data-supported time interval, suggesting more robust extrapolation capabilities.

The residuals of both models remain bounded and exhibit no systematic late-time drift, showing that neither ROM loses the constrained energy-balance structure during extrapolation. LaSDI thus retains qualitative consistency with the equation, while POD--PINN satisfies it substantially more accurately and uniformly. 
%
%
%
%
%

%
%
%
%
%
\section{Conclusion}
%
%
This work compares two physics-guided parametric ROMs for Rayleigh--Taylor mixing across the transition to turbulence and the self-similar regime. POD--PINN combines an interpretable linear spatial decomposition with nonlinear neural-network dynamics, whereas LaSDI constructs a nonlinear latent representation governed by a linear evolution law. Their comparison therefore isolates a central modeling trade-off: whether complexity should be carried by the reduced representation or by the dynamics evolving within it.

For the POD--PINN approach, the separation between spatial modes and temporal amplitudes preserves one of the main advantages of classical modal decompositions. The leading POD modes can be directly associated with the global evolution and self-similar structure of the profiles, whereas the secondary modes describe the transient corrections associated with the development of coherent structures and the transition to turbulence. Symmetry, realizability, and self-similar scaling are introduced directly into the construction of the reduced representation, providing physically meaningful coordinates and facilitating the interpretation of the reconstructed fields. This spatial interpretability is obtained at the cost of a less interpretable temporal model. Unlike a classical POD--Galerkin formulation, the evolution of the modal amplitudes is not governed by an explicit low-dimensional system whose individual interaction terms can be analyzed directly, but by a neural network. The resulting closure is therefore implicit. 

In contrast to the static solution map approach of POD-PINN, the second approach adopts a dynamical evolution model. Next, motivated by the observation that the temporal extent of observed trajectories in spatially extended dynamical systems is typically much too short, the second approach focuses simultaneously on keeping the latent space dimension very low and the dynamics simple. Limiting dynamics to be linear then requires the coordinate transformation step to be nonlinear to be able to limit the latent space dimension to be very low. 
The present results show that a latent dimension of only 4 (with minor degradation in performance at 2) is sufficient to obtain an efficient representation of the considered profiles while retaining linear latent dynamics. To reiterate, we find it surprising that strongly nonlinear and transient regimes can be reasonably well represented by linear dynamics in a very low dimensional latent space. Although the latent coordinates do not admit the direct physical interpretation of POD modes, the linear evolution law offers another route toward interpretability. In particular, classical dynamical-systems tools can be applied to the identified latent operator. The analysis of its eigenvalues and eigenvectors could reveal characteristic time scales, stability properties, bifurcation, dominant latent directions and the fixed points associated with the self-similar regime. Establishing explicit links between these latent structures and physical observables therefore constitutes a promising perspective.



Overall, the results demonstrate that the choice between linear and nonlinear reduced representations should not be based solely on reconstruction accuracy. POD--PINN is particularly suitable when physically identifiable spatial structures and explicit control over conservation properties are required. LaSDI is especially attractive when strong a priori assumptions about the relevant physics are unavailable: beyond providing a compact representation and efficient long-time integration, its learned latent structure and linear dynamics can serve as a data-driven discovery tool for formulating hypotheses about the underlying organization of the dynamics.

We experimented further with the second approach to see if prediction accuracy could be further improved by allowing for nonlinearity in the dynamics. This seemed to suggest that obtaining significant improvements over comparable accuracy of the two approaches presented may not be possible. As such, we speculate that nonlinearity in one of the two steps---coordinate transformation and latent-space dynamics/solution map---may be sufficient. This reinforces and reiterates our view that the two approaches are complementary rather than competing. Future work could combine their respective strengths, for example by introducing explicit physical observables into the LaSDI latent space, analyzing the spectrum of its latent dynamics matrix, or replacing the black-box temporal component of POD--PINN with a partially interpretable learned dynamical system. Such developments would contribute to reduced-order models that are simultaneously compact, predictive, physically consistent, and informative for the analysis and modeling of turbulent Rayleigh--Taylor mixing.
\newpage
\appendix
\section{Linear encoding-decoding based surrogate}
\subsection{POD details}
\label{app:POD_details}
This appendix provides additional details on the POD representation introduced in Section~\ref{sec2}. It first establishes the relation between the modal amplitudes and the zero-dimensional quantities, then clarifies the connection between the leading POD modes and the self-similar profiles, and finally separates the error associated with the truncated spatial basis from that introduced by the PINN prediction of the modal amplitudes.
\begin{figure}[H]
    \centering
    \includegraphics[width=0.9\textwidth]{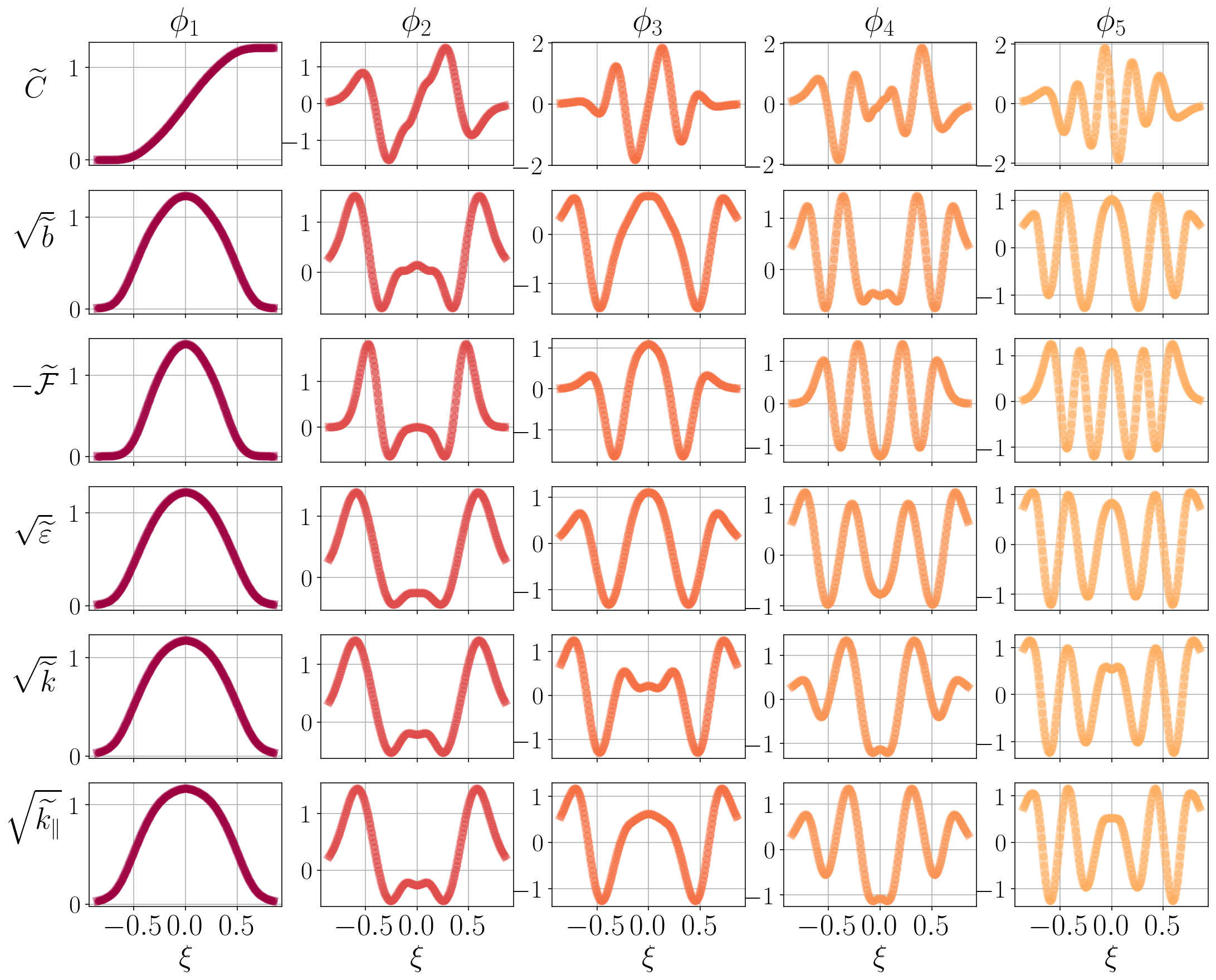}
    \caption{First five POD modes of the self-similar profiles $\widetilde{C}$, $\widetilde{b}$, $-\widetilde{\mathcal{F}}$, $\widetilde{\varepsilon}$, $\widetilde{k}$, and $\widetilde{k}_{\parallel}$. Non-negative quantities are decomposed through their square roots. Five modes are retained for each quantity, except for $\widetilde{C}$, for which four are used.}
    \label{fig:modes_comparaison_1_a_5_symm}
\end{figure}
\subsubsection{Zero-dimensional quantities}\label{app:0D_Q_POD}

The zero-dimensional quantities defined in Eq.~\eqref{eq:zero_dimensional_quantity} can be evaluated directly from the reduced coordinates, without reconstructing the profiles on the spatial grid.

\begin{equation}
\renewcommand{\arraystretch}{2.2}
\begin{array}{c|c|c|c|c|c}
\widetilde{Q}
& \widetilde{k}
& \widetilde{k}_{\parallel}
& \widetilde{\varepsilon}
& \widetilde{b}
& -\widetilde{\mathcal{F}}
\\
\hline
Q_{0\mathcal{D}}(t)
&
\displaystyle
L(t)\sum_{i=1}^{r=5}
\left(a_i^{\widetilde{k}}\right)^2
&
\displaystyle
L(t)\sum_{i=1}^{r=5}
\left(a_i^{\widetilde{k}_{\parallel}}\right)^2
&
\displaystyle
\sqrt{L(t)}\sum_{i=1}^{r=5}
\left(a_i^{\widetilde{\varepsilon}}\right)^2
&
\displaystyle
\sum_{i=1}^{r=5}
\left(a_i^{\widetilde{b}}\right)^2
&
\displaystyle
\sqrt{L(t)}\sum_{i=1}^{r=5}
m_i^{\widetilde{\mathcal{F}}}
a_i^{\widetilde{\mathcal{F}}}
\end{array}
\label{eq:POD_0D_quantity}
\end{equation}

For the positive quantities represented through their square roots, orthonormality eliminates the cross terms in the spatial integral. The spatial moment of the $i$th mass-flux mode is defined as

\begin{equation}
m_i^{\widetilde{\mathcal{F}}}
=
\int_{\Omega}
\phi_i^{\widetilde{\mathcal{F}}}(\xi)
\,d\xi .
\end{equation}
where $\Omega=[-0.85,0.85]$. This interval is wider than the nominal mixing-layer region $[-0.5,0.5]$ because the turbulent profiles develop diffuse tails that extend beyond its boundaries. Retaining these tails improves the accuracy of both the modal representation and the spatially integrated quantities.

Figure~\ref{fig:modes_comparaison_1_a_5_symm} presents the first five modes obtained after applying the symmetry and realizability transformations described in Section~\ref{sec2}. The leading modes are smooth and reproduce the dominant spatial support of the mixing layer. The secondary modes contain an increasing number of extrema and progressively finer spatial scales, consistently with their role in representing transient profile deformations. Five modes are retained for all quantities except the mean concentration, for which the first four modes are sufficient; its fifth mode is nevertheless displayed for comparison.
\begin{figure}[h]
    \centering
    \begin{minipage}[c]{0.49\textwidth}
        \centering
        \includegraphics[width=\textwidth]{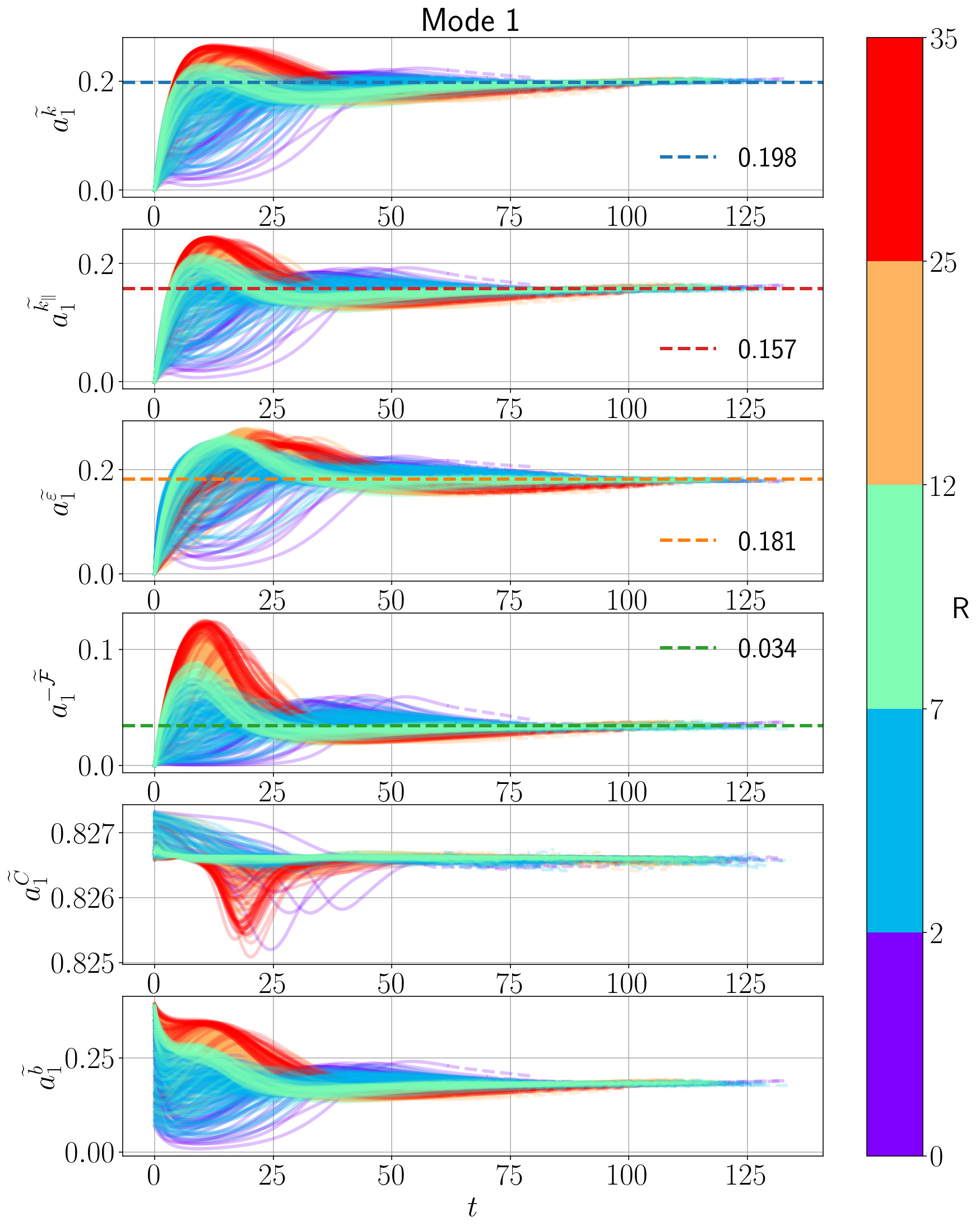}
    \end{minipage}%
    \hfill
    \begin{minipage}[c]{0.49\textwidth}
        \centering
        \includegraphics[width=\textwidth]{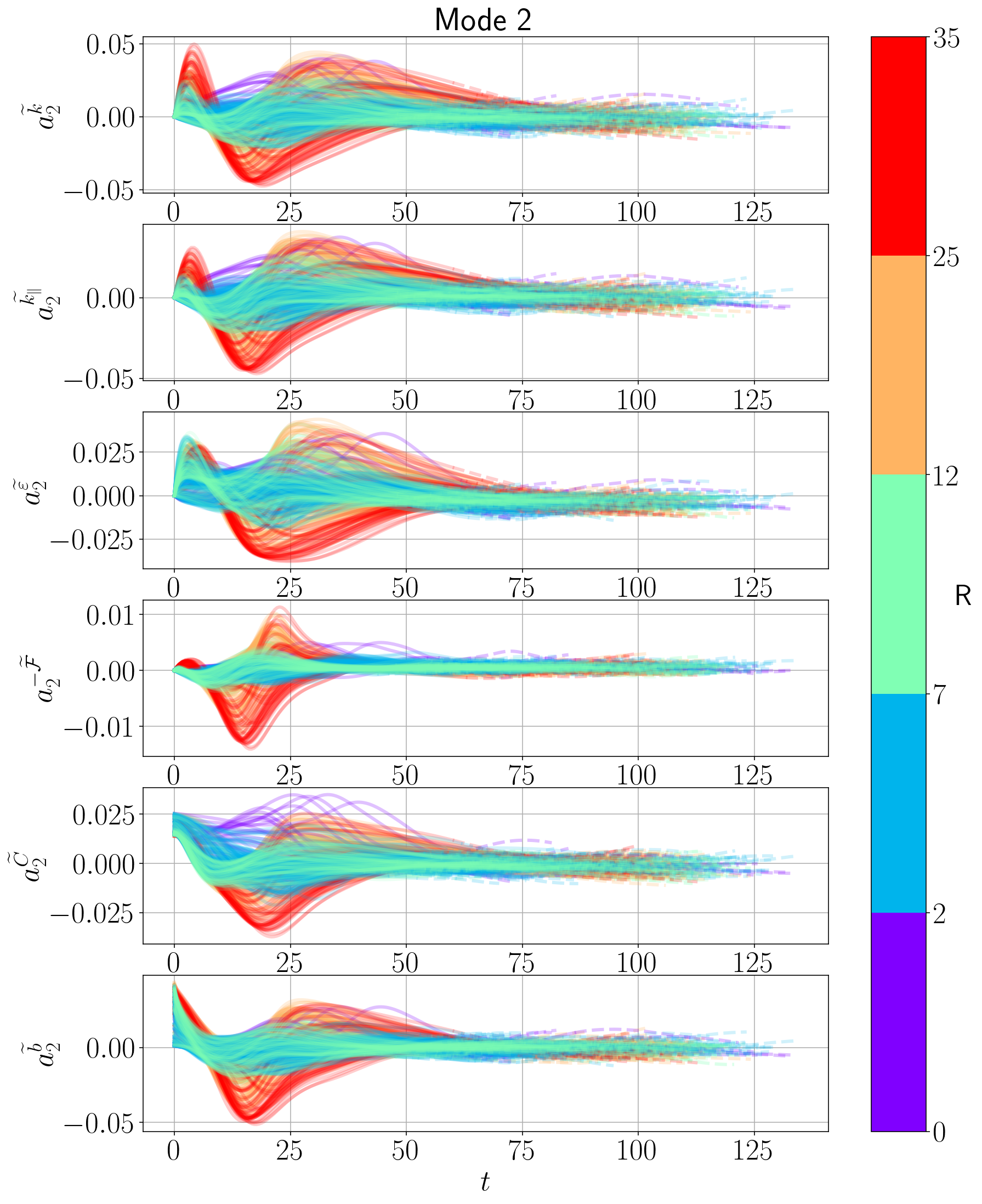}
    \end{minipage}
    \caption{DNS-projected temporal amplitudes of the first POD mode (left) and second POD mode (right) for all modeled quantities. Each curve corresponds to one trajectory and is colored according to $\mathsf{R}$. The dashed segments represent the final $25\%$ of each trajectory, which are excluded from the training of the PINN and reserved for temporal-extrapolation assessment.}
    \label{fig:amp_pod_mode_1_2}
\end{figure}

\subsubsection{Asymptotic interpretation of the leading POD modes}\label{app:interpret_POD_modes}

Because the POD is performed directly on the self-similar quantities, higher-order amplitudes decay toward zero. The leading amplitude approaches a constant, $\gamma_Q$, and the asymptotic profile is governed by the first mode. For all non-negative quantities represented through their square roots,
\begin{equation}
\begin{gathered}
\lim_{t\rightarrow\infty}
\sqrt{\widetilde{Q}(\xi,t)}
\approx
\phi_1^{\widetilde{Q}}(\xi)\,\gamma_{\widetilde{Q}}
=
\sqrt{f_Q(\xi)},\\[0.3em]
\gamma_{\widetilde{Q}}
=
\lim_{t\rightarrow\infty}
a_1^{\widetilde{Q}}(t),
\qquad
\widetilde{Q}
\in
\left\{
\widetilde{k},
\widetilde{k}_{\parallel},
\widetilde{\varepsilon},
\widetilde{b}
\right\}.
\end{gathered}
\label{eq:positive_ss_pod}
\end{equation}
For the mean concentration and the modeled negative mass flux, which are decomposed without the square-root transformation, the relation remains linear:
\begin{equation}
\begin{gathered}
\displaystyle
\lim_{t\rightarrow\infty}
\widetilde{Q}(\xi,t)
\approx
\phi_1^{\widetilde{Q}}(\xi)\,\gamma_{\widetilde{Q}}
=
f_Q(\xi)
\\[0.5em]
\displaystyle
\gamma_{\widetilde{Q}}
=
\lim_{t\rightarrow\infty}
a_1^{\widetilde{Q}}(t),
\qquad
\widetilde{Q}
\in
\left\{
\widetilde{C},
-\widetilde{\mathcal{F}}
\right\}.
\end{gathered}
\label{eq:ss_pod}
\end{equation}
The leading POD mode is therefore directly proportional to the asymptotic self-similar profile for $\widetilde{C}$ and $-\widetilde{\mathcal{F}}$, and to its square root for the positive quantities.

%
\subsection{PINN implementation details}
\label{app:PINN}
%
\noindent
This appendix complements the POD--PINN presentation in Sec.~\ref{sec2} by collecting the implementation details omitted from the main text. It describes the input and output preprocessing, the complete loss function, the construction of the physical residuals, the collocation-point strategy, and the reconstruction-error assessment.

\subsubsection{Preprocessing and output parametrization}

For both POD--PINN and LaSDI, only the first $75\%$ of the temporal samples from each training trajectory are used for model fitting, while the final $25\%$ are withheld to assess temporal-extrapolation capabilities, as illustrated in Fig.~\ref{fig:amp_pod_mode_1_2}.

The four initial-condition parameters are logarithmically transformed, while the dimensionless time is left unchanged:

\begin{equation}
    \mathbf{x_{log}} =
    \left(
    \log(\mathsf{R}),
    \log(\mathsf{B}),
    \log(\mathsf{S}),
    \log(\mathsf{D}),
    t
    \right)^T .
\end{equation}

All outputs, including $L$, are subsequently normalized to $[0,1]$. A squared Softplus transformation is applied to the $L$ output to promote its expected quadratic late-time growth, see Eq.\eqref{eq:self_similar_mixing_width}. Softplus activations are also used in the hidden layers because their smoothness facilitates derivative evaluation by automatic differentiation.
\subsubsection{Data-driven loss function}\label{app:loss_data}

Terms of the loss function are constructed from mean-squared-error (MSE). For two tensors $\hat{\mathbf{y}}$ and $\mathbf{y}$ of identical shape, the MSE is defined as
\begin{equation}
	\mathrm{MSE}(\hat{\mathbf{y}}, \mathbf{y}) = \frac{1}{n} \sum_{j=1}^n \left( \hat{y}_j - y_j \right)^2,
	\label{eq:MSE}
\end{equation}
where $n$ is the total number of elements in each tensor and $j$ indexes their flattened entries. The data-driven loss is then written as the weighted sum of four contributions:
\begin{equation}
    \mathcal{L}^{Data}
    =
    \mathcal{L}_{L}
    +
    \mathcal{L}_{L_0}
    +
    \mathcal{L}_{\dot{L}}
    +
    \mathcal{L}_{pod}.
\end{equation}

The initial value $L_0=L(t=0)$ is penalized separately to prevent it from being underweighted relative to the remaining time samples:
\begin{equation}
    \mathcal{L}_{L}
    =
    \lambda_{L}
    \mathrm{MSE}
    (\hat{L}_{t\neq0},L_{t\neq0}),
    \qquad
    \lambda_L=5 ,
    \label{eq:loss_L_star}
\end{equation}
\begin{equation}
    \mathcal{L}_{L_0}
    =
    \lambda_{L_0}
    \mathrm{MSE}
    (\hat{L}_0,L_0),
    \qquad
    \lambda_{L_0}=50 .
    \label{eq:loss_L0_star}
\end{equation}

\begin{figure}[H]
    \centering
    \includegraphics[width=1\textwidth]{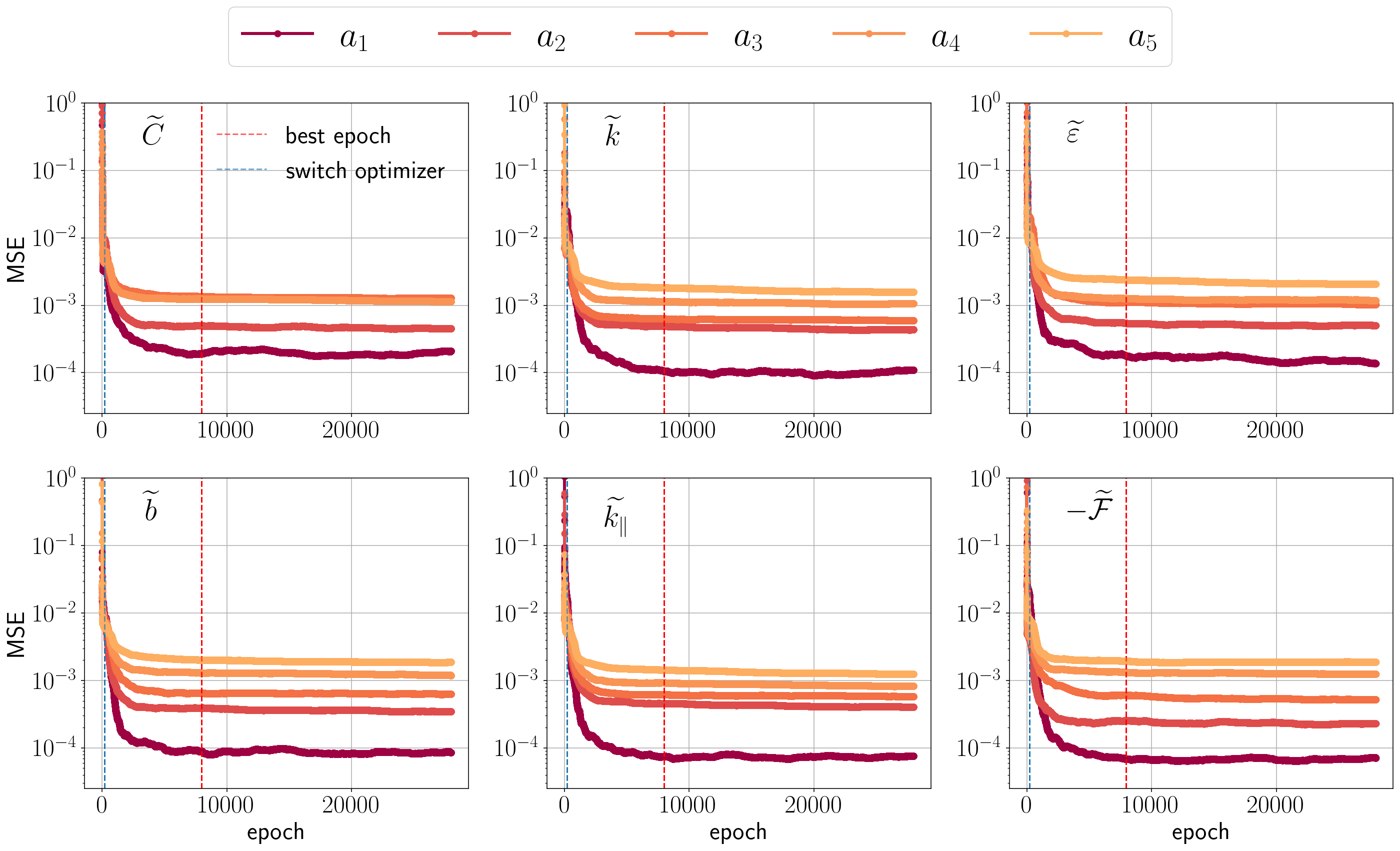}
    \caption{\label{fig:loss_amp_pod} 
    Mean squared error of each POD amplitude for each quantity during the training of the PINN on the training set. The red dashed line indicates the epoch at which the PINN achieves its best performance on the validation set. The blue dashed line at 200 epochs marks the switch of the optimizer from Adam to SSBroyden.}
\end{figure}

The expansion-rate term compares the derivative obtained by automatic differentiation with the DNS reference:
\begin{equation}
    \mathcal{L}_{\dot{L}}
    =
    \lambda_{\dot{L}}
    \mathrm{MSE}
    (\partial_{t}\hat{L},\dot{L}),
    \qquad
    \lambda_{\dot{L}}=15\times10^2 .
    \label{eq:loss_dtL_star}
\end{equation}
The modal term uniformly weights the $n_{pod}=29$ rescaled POD amplitudes; energy-based weighting was also tested but did not improve the prediction accuracy:
\begin{equation}
    \mathcal{L}_{pod}
    =
    \lambda_{pod}
    \frac{1}{n_{pod}}
    \sum_{\widetilde{Q}}
    \mathrm{MSE}
    \left(
        \hat{a}_i^{\widetilde{Q}},
        a_i^{\widetilde{Q}}
    \right),
    \qquad
    n_{pod}=29,
    \qquad
    \lambda_{pod}=0.5 .
    \label{eq:loss_pod}
\end{equation}
Figure~\ref{fig:loss_amp_pod} shows that the learned errors follow the modal hierarchy, with the leading modes predicted most accurately.
%
%
%
%
%
\subsubsection{Physics-informed loss function}\label{app:loss_physics}

The physics-informed contribution used during training is
\begin{equation}
	\mathcal{L}^{PI}
	=
	\mathcal{L}_{\dot{L}>0}
	+
	\mathcal{L}_{k_{0\mathcal{D}}}
	+
	\mathcal{L}_{C_{1\mathcal{D}}},
\end{equation}

Negative predicted values of $\partial_t\hat{L}$ are penalized by
\begin{equation}
	\mathcal{L}_{\dot{L}>0}
	=
	\lambda_{\dot{L}>0}
	\,
	\mathrm{MSE}
	\left(
	\mathbf{0},
	\min
	\left(
	0,
	\partial_{t}\hat{L}
	\right)
	\right),
	\qquad
	\lambda_{\dot{L}>0}=10^6.
\end{equation}

The spatially integrated balance of kinetic energy is
\begin{equation}
	d_{t} K
	+
	\frac{\dot{L}}{L}
	K
	+
	2 F 
	+
	E
	=
	0.
\label{eq:K_0D_eq}
\end{equation}
Substitution of the retained POD expansions \eqref{eq:POD_0D_quantity}, gives
\begin{equation}
	Ld_{t} a_i^{\widetilde{k}^{2}}
	+
	2
	a_i^{\widetilde{k}^{2}}
    \dot{L}
	-
	2m_i^{\widetilde{\mathcal{F}}}
	a_i^{\widetilde{\mathcal{F}}}
    \sqrt{L}
	+
	a_i^{\widetilde{\varepsilon}^{2}}
    \sqrt{L}
	=
	r_k,
\label{eq:K_0D_eq_ROM}
\end{equation}
where $m_i^{\mathcal{F}}$ is the spatial integral of mode $i$ and repeated indices are summed. The truncation residual is denoted $r_k^{\mathrm{pod}}$ for DNS-projected amplitudes and $r_k^{\mathrm{pinn}}$ for PINN predictions. Only the latter is minimized:
\begin{equation}
	\mathcal{L}_{k_{0\mathcal D}}
	=
	\lambda_{k_{0\mathcal D}}
	\,
	\mathrm{MSE}
	\left(
	\mathbf{0},
	r_k^{\mathrm{pinn}}
	\right),
	\qquad
	\lambda_{k_{0\mathcal D}}=0.5.
\label{eq:loss_eq_K0D}
\end{equation}
This formulation lets the learned retained-mode dynamics accounts implicitly for unresolved contributions, without fitting the DNS residual outside the available sampling points.

The governing advection--diffusion equation is
\begin{equation}
	\partial_{t} \overline{C}(z,t)
	+
	\partial_z \mathcal{F}(z,t)
	-
	\partial_{zz} \overline{C}(z,t)
	=
	0.
\end{equation}
In the self-similar formulation,
\begin{equation}
	 \partial_{t} \widetilde{C}(\xi, t)
	-\xi \frac{\dot{L}}{L} \partial_\xi \widetilde{C}(\xi, t)
	+
	\frac{1}{\sqrt{L}} \partial_\xi \widetilde{\mathcal{F}}(\xi, t)
	- \frac{1}{L^{2}}\partial_{\xi \xi} \widetilde{C}(\xi, t)
	=0.
    \label{eq:self_sim_eq_C}
\end{equation}
Substitution of the retained POD expansions yields
\begin{equation}
	\phi_i^{\widetilde{C}}
	d_{t} a_i^{\widetilde{C}}
	-
	\xi
	\frac{\dot{L}}{L}
	a_i^{\widetilde{C}}
	d_\xi \phi_i^{\widetilde{C}}
	-
	\frac{a_i^{\widetilde{\mathcal{F}}}}
	     {\sqrt{L}}
	d_\xi \phi_i^{\widetilde{\mathcal{F}}}
	-
	\frac{a_i^{\widetilde{C}}}
	     {L^{2}}
	d_{\xi\xi}\phi_i^{\widetilde{C}}
	=
	r_C (\xi,t),
\end{equation}
where the explicit dependence on $(\xi,t)$ is omitted. The corresponding loss is
\begin{equation}
	\mathcal{L}_{C_{1\mathcal{D}}}
	=
	\lambda_{C_{1\mathcal{D}}}
	\,
	\mathrm{MSE}
	\left(
	\mathbf{0},
	r_C^{\mathrm{pinn}}
	\right),
	\qquad
	\lambda_{C_{1\mathcal{D}}}=10^2.
\label{eq:loss_eq_C0D}
\end{equation}
Because $r_C^{\mathrm{pinn}}$ explicitly involves the POD modes and their spatial derivatives, this term constrains both the modal dynamics and the reconstructed profile shapes while providing an implicit closure for discarded modes.
%
%
%
\subsubsection{Collocation points and residual-based attention}\label{app:colloc_pts}
%

The physical residuals are evaluated at collocation points, which require no reference DNS values and can therefore extend beyond the available trajectories. To provide extrapolation constraints without sampling unphysical configurations, these points are generated by Latin Hypercube Sampling \citep{Stein_1987} within
\begin{equation}
	t \in [1,200],
	\qquad
	\mathsf{R} \in [0.2,35],
	\qquad
	\mathsf{B} \in [0.01,1.7],
	\qquad
	\mathsf{S} \in [0.5,10],
	\qquad
	\mathsf{D} \in [0.7,20].
\label{eq:domaine_colloc_pts}
\end{equation}
This domain covers both the DNS database and neighboring unexplored regions. During training, Residual-Based Attention \citep{Anagnostopoulos_2024} adaptively increases the weight of points with large physical residuals, thereby concentrating the optimization on the most difficult temporal and parametric regimes.

%
%
%
%
%
%
%
%
%
\subsubsection{Reconstruction accuracy}\label{app:reconstruct_pod_vs_pinn}

\begin{figure}[H]
    \centering
    \includegraphics[width=1\textwidth]{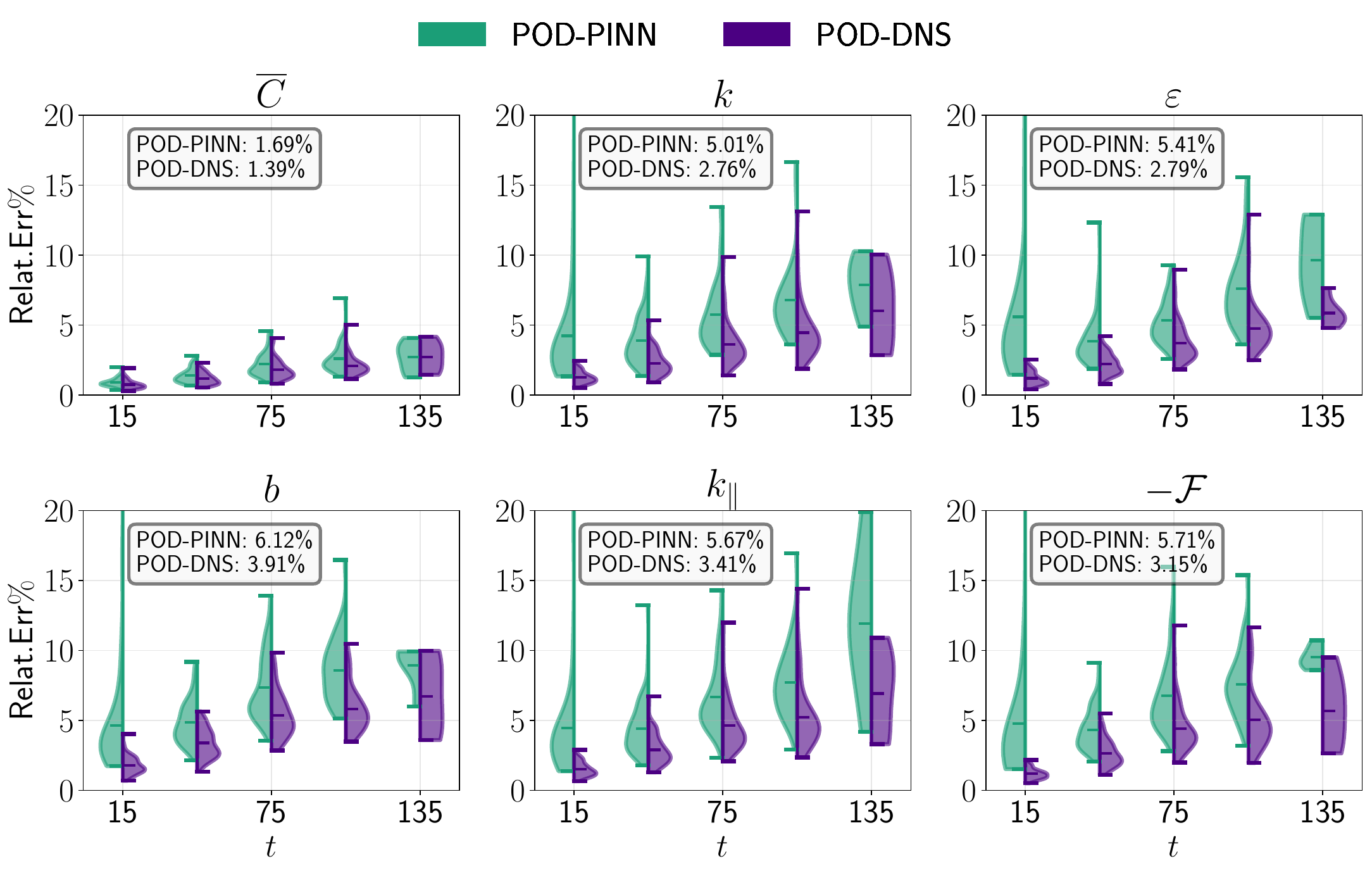}
    \caption{Temporal distributions of the mean relative profile-reconstruction errors with respect to the DNS for the POD--PINN surrogate (green) and the direct POD reconstruction using DNS-projected amplitudes (purple). For each test trajectory and time interval, the relative error is averaged first over $\xi\in[-0.5,0.5]$ and then over the snapshots contained in that interval; the resulting trajectory-wise errors form the violin distributions. Restricting the spatial interval avoids artificially large relative errors near profile boundaries where the reference values approach zero. Because the DNS trajectories do not all reach the same final time, the latest intervals contain fewer samples. The values displayed in each panel are the mean errors over the complete test set.}
    \label{fig:relat_err_pod_time}
\end{figure}

Finally, Fig.~\ref{fig:relat_err_pod_time} separates the error caused by POD truncation from that introduced by the learned temporal dynamics. The POD--DNS reconstruction uses amplitudes obtained by projecting the DNS profiles onto the truncated basis and therefore isolates the spatial-representation error. The POD--PINN reconstruction instead uses the predicted amplitudes, so that the difference between the two distributions measures the additional error introduced by the neural network.

The direct POD reconstruction provides the lower-error baseline for all quantities, as expected, while the POD--PINN distributions follow the same temporal trends with an additional contribution from the predicted amplitudes. The errors generally increase and broaden with time, but the mean POD--PINN error over the complete test set remains between approximately $2\%$ and $6\%$, depending on the quantity. The concentration profile is reconstructed most accurately, whereas the square-root-transformed turbulent quantities are more sensitive to errors in the predicted amplitudes after transformation back to physical space. Overall, the selected basis remains sufficiently compact to preserve physical interpretability while providing an accurate representation for the surrogate model.
%
%
%
%
%
\section{Additional ROM comparisons}
\label{app:additional_rom_comparisons}

\begin{figure}[p]
    \centering
    \includegraphics[width=0.9\textwidth]{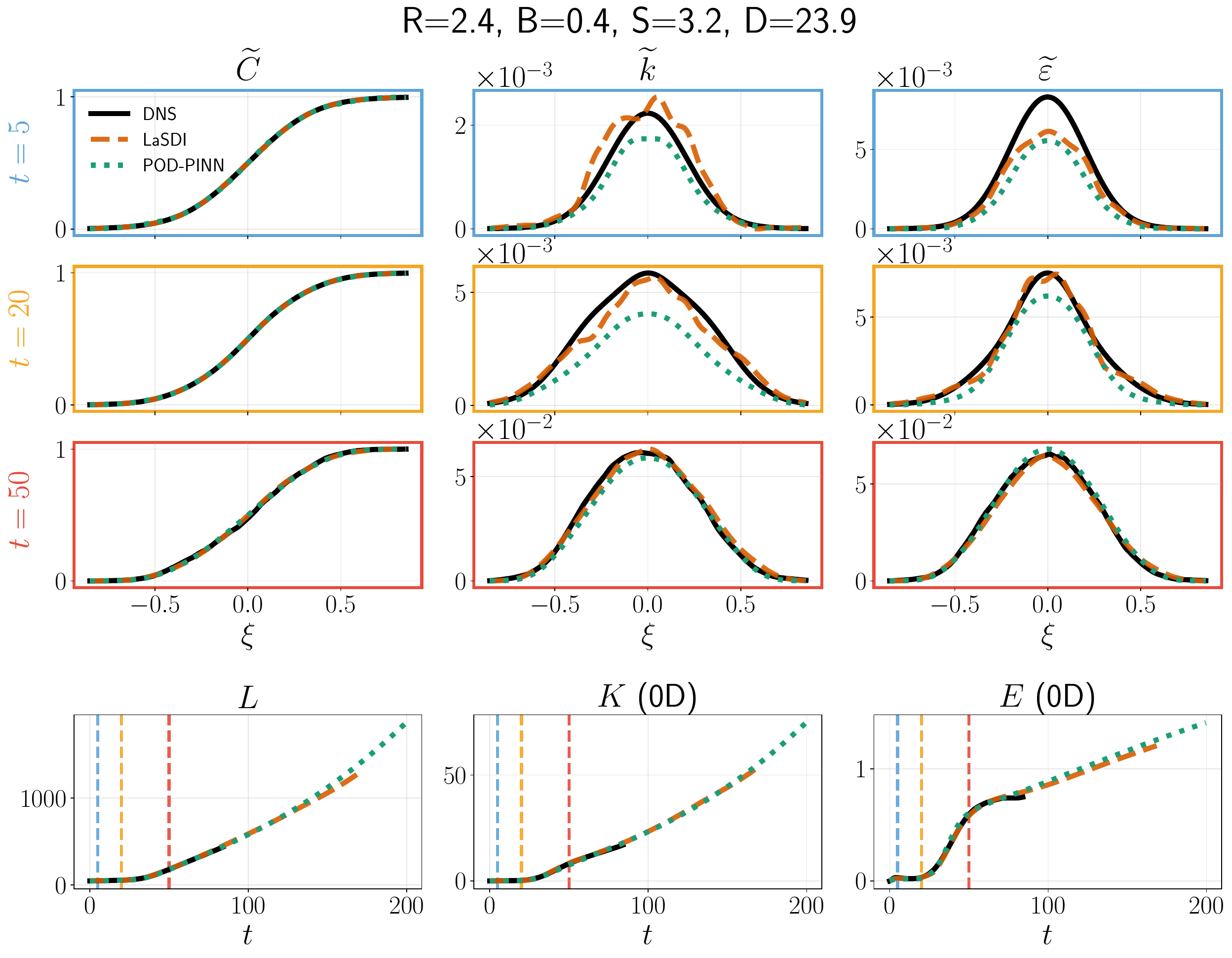}
    \caption{Detailed comparison of the DNS, LaSDI, and POD-PINN predictions for the representative diffusion-dominated trajectory shown in purple in Fig.~\ref{fig:surrogate_comparison_0D_phase}. Instantaneous profiles are shown together with the evolution of the corresponding zero-dimensional quantities.}
    \label{fig:compare_profiles_diffusive}
\end{figure}

\begin{figure}[p]
    \centering
    \includegraphics[width=0.9\textwidth]{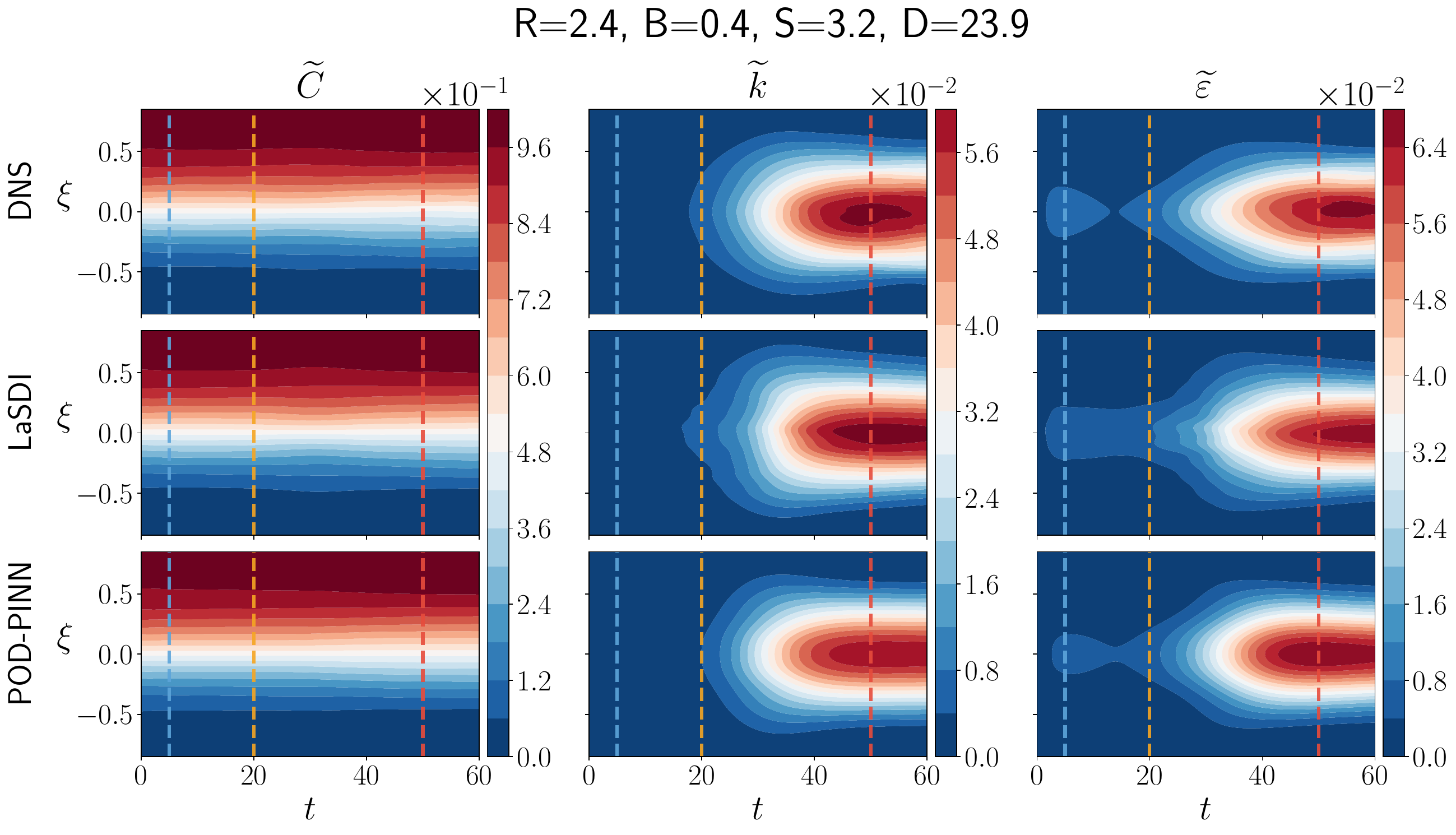}
    \caption{Space--time evolution of the self-similar profiles for the representative diffusion-dominated trajectory shown in purple in Fig.~\ref{fig:surrogate_comparison_0D_phase}, comparing the DNS reference with the LaSDI and POD-PINN predictions during the transition to turbulence.}
    \label{fig:compare_profiles_stack_diffusive}
\end{figure}

In this section, we provide two additional comparisons between the DNS and surrogate-model predictions for representative test cases shown in the phase diagram of Fig.~\ref{fig:surrogate_comparison_0D_phase}.
For the diffusion-dominated trajectory with initial Reynolds number $\mathsf{R}=2.4$, selected one-dimensional profiles of the mean concentration $\widetilde{C}$, kinetic energy $\widetilde{k}$, and dissipation $\widetilde{\varepsilon}$ are presented in Fig.~\ref{fig:compare_profiles_diffusive}. The corresponding space--time evolution is shown in Fig.~\ref{fig:compare_profiles_stack_diffusive}.

\begin{figure}[p]
    \centering
    \includegraphics[width=0.9\textwidth]{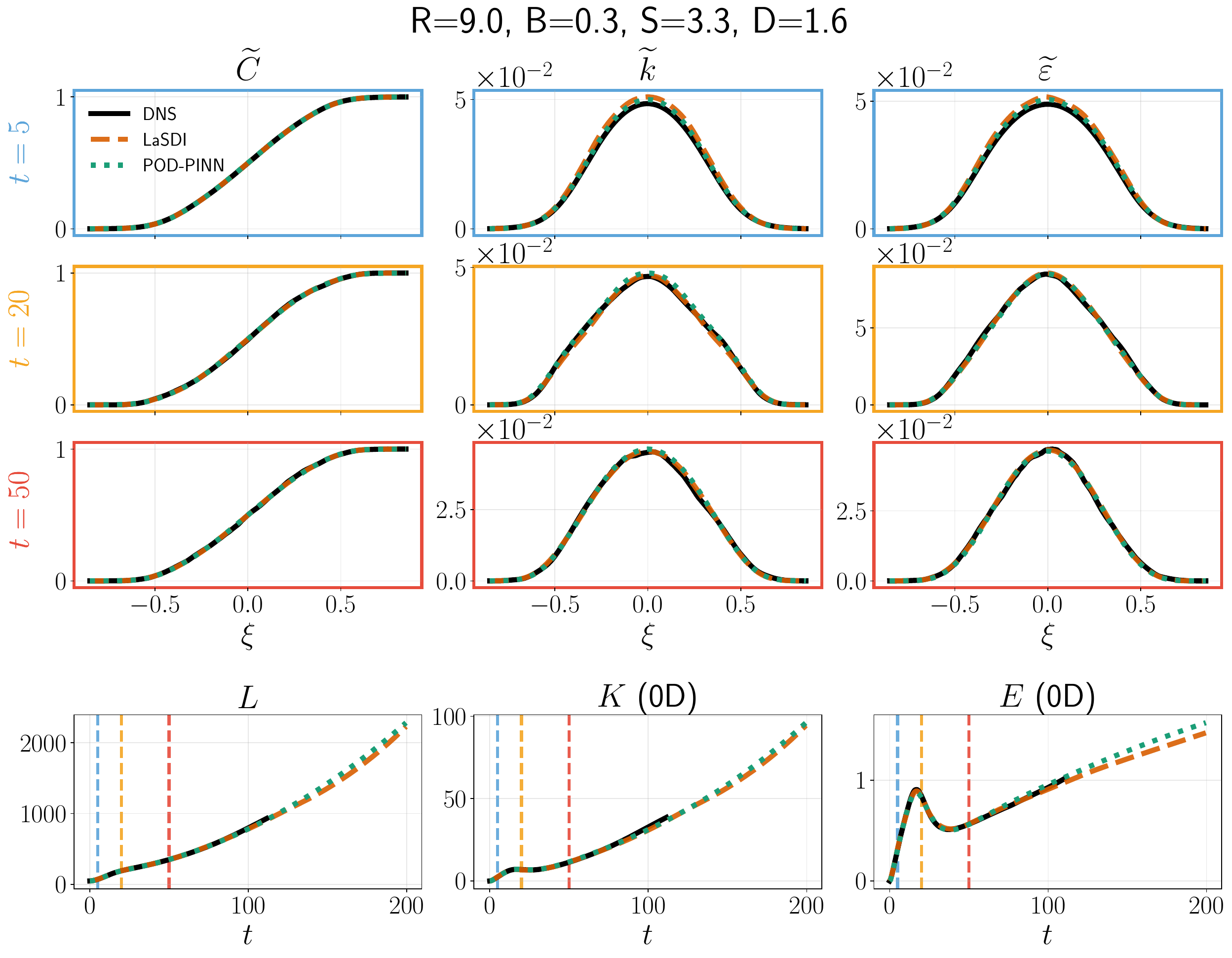}
    \caption{Detailed comparison of the DNS, LaSDI, and POD-PINN predictions for the representative intermediate trajectory shown in cyan in Fig.~\ref{fig:surrogate_comparison_0D_phase}. Instantaneous profiles are shown together with the evolution of the corresponding zero-dimensional quantities.}
    \label{fig:compare_profiles_intermediate}
\end{figure}
\begin{figure}[p]
    \centering
    \includegraphics[width=0.9\textwidth]{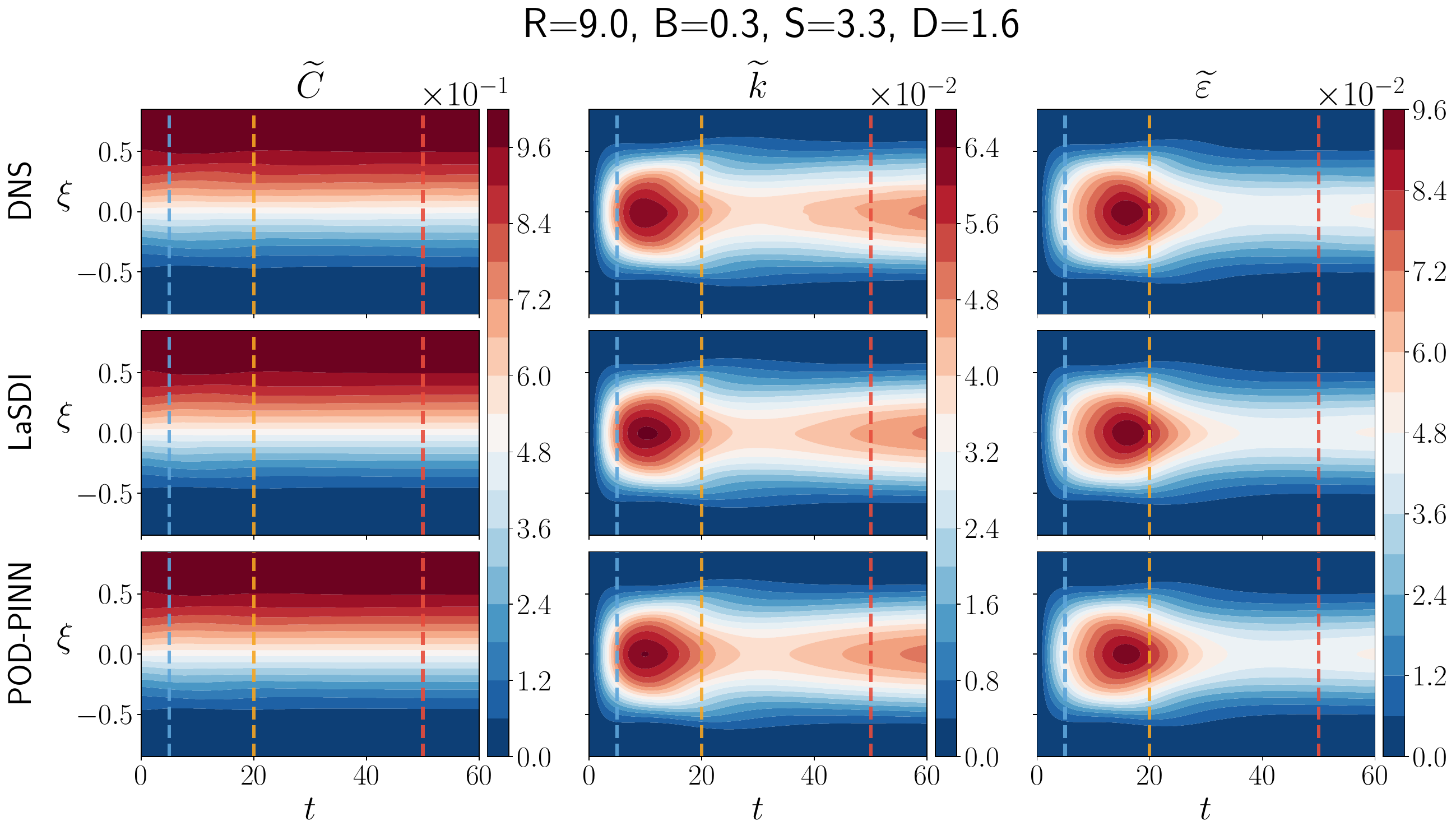}
    \caption{Space--time evolution of the self-similar profiles for the representative intermediate trajectory shown in cyan in Fig.~\ref{fig:surrogate_comparison_0D_phase}, comparing the DNS reference with the LaSDI and POD-PINN predictions during the transition to turbulence.}
    \label{fig:compare_profiles_stack_intermediate}
\end{figure}
Similarly, Fig.~\ref{fig:compare_profiles_intermediate} presents the one-dimensional profiles and corresponding zero-dimensional quantities for a representative trajectory with an intermediate Reynolds number of $\mathsf{R}=9$, while Fig.~\ref{fig:compare_profiles_stack_intermediate} shows the associated space--time evolution.

\section{Total Energy conservation}
\label{annex:energy_conservation_section}

Under the Boussinesq approximation, the dimensionless density field, normalized by the average of the pure-fluid densities, is related to the heavy-fluid concentration field by
\begin{equation}
    \rho(\boldsymbol{x},t)
    =
    1 + 2\mathcal{A} \left( C(\boldsymbol{x},t) - \frac{1}{2} \right).
\end{equation}

Assuming that the mean concentration profile is piecewise linear, the variation in potential energy relative to the initial condition, $\Delta \mathcal E _p$, can then be estimated as \citep{Cook_2004}
\begin{equation}
    \Delta\mathcal{E}_p(t)
    =
    \frac{1}{\mathcal{V}}
    \int_{\mathcal{V}}
    \left[\rho(\boldsymbol{x},t)-\rho(\boldsymbol{x},0)\right]
    z\,\mathrm{d}\boldsymbol{x}
    =
    -\frac{1}{12}L(t),
    \label{eq:potential_energy}
\end{equation}
where $\mathcal{V}$ denotes the fluid volume (again quantities are non dimensionalized). Thus, the magnitude of the potential energy released into the system increases as the mixing layer grows.

The total-energy balance can then be written as
\begin{equation}
\left(
\frac{d}{dt}
+
\frac{\dot{L}}{L}
\right)
\left(
K
+
\Delta\mathcal{E}_p
\right)
=
-E,
\label{eq:energy_conservation}
\end{equation}
Substituting Eq.~\eqref{eq:potential_energy} into Eq.~\eqref{eq:energy_conservation} gives
\begin{equation}
\frac{dK}{dt}
+
\frac{\dot{L}}{L}
K
-
\frac{\dot{L}}{6}
+
E
=
0.
\label{eq:energy_conservation_2}
\end{equation}
The ratio of turbulent kinetic energy to the magnitude of the released potential energy is then defined as
\begin{equation}
    \Pi
    =
    \frac{K}{\lvert\Delta\mathcal{E}_p\rvert}
    =
    12\frac{K}{L}.
\end{equation}
This dimensionless ratio is physically bounded between 0 and 1:
\begin{equation}
0 \leq \Pi \leq 1.
\end{equation}
\clearpage
\bibliographystyle{plainnat}   
\bibliography{bibliography}

\end{document}